\documentclass[aps,showpacs,twocolumn,longbibliography,secnumarabic,superscriptaddress]{revtex4-2}
\usepackage{graphicx}
\usepackage{dcolumn}
\usepackage{bm}
\usepackage{braket}
\usepackage{leftidx}
\usepackage{cancel}
\usepackage{amsmath}
\usepackage{epstopdf}
\usepackage{color}
\usepackage[mathcal]{eucal}
\usepackage{float}
\usepackage{amssymb}
\usepackage{yfonts}
\usepackage{color}

\usepackage{scalerel}

\usepackage{pdfpages}
\makeatletter
\AtBeginDocument{\let\LS@rot\@undefined}
\makeatother

\begin{document}
\date{\today}
\title{Observable criterion for collective entanglement in Boson-lattice system}

\author{Mehmet Emre Tasgin} \thanks{metasgin@hacettepe.edu.tr }
\affiliation{Institute of Nuclear Sciences, Hacettepe University, 06800, Ankara, Turkey}
\affiliation{Department of Physics, Texas A \& M University at Qatar, POBox 23874, Doha, Qatar}

\author{Hyunchul Nha} \thanks{hyunchul.nha@qatar.tamu.edu}
\affiliation{Department of Physics, Texas A \& M University at Qatar, POBox 23874, Doha, Qatar}
\affiliation{School of Computational Sciences, Korea Institute of Advanced Study, Seoul 02455, Korea}

\begin{abstract}
	An optical lattice with cold trapped atoms represents a quantum system of fundamental importance as it enables the study of quantum many-body system in a controllable way. It is thus necessary to develop theoretical and experimental tools to explore quantum correlation in such systems to advance our understanding of many-body physics. While previous works have identified some profound aspects of quantum entanglement using e.g. entanglement entropy,  there exists a critical demand to have an experimentally accessible tool to investigate many-body quantum entanglement in a broad context. We present an entanglement criterion characterizing collective entanglement in Boson lattice systems and enabling experimental observation readily. On applying our approach to the extended Bose-Hubbard model, we show that our criterion witnesses phase transitions such as Mott insulator--superfluid and Mott insulator--charge density wave transitions. Remarkably, it also makes it possible to detect multipartite entanglement among boson lattice sites in a rigorous sense. Our criterion can be experimentally tested via Raman scattering or time-of-flight methods, which are within the reach of current technology. 
\end{abstract}

\maketitle

{\small \bf Introduction}---
Cold atoms trapped in optical lattice provide a versatile platform for quantum simulation to study complex physical phenomena in quantum many-body systems with remarkable experimental controllability \footnote{C. Gross and W. S. Bakr, Quantum gas microscopy for single atom and spin detection, Nature Phys. {\bf 17}, 1316 (2021).}. Unlike rigid crystals, an optical lattice allows a continuous tuning of parameters like interatomic interactions, lattice depth and size~\cite{inouye1998observation,theis2004tuning,schafer2020tools} enabling a comprehensive study of many-body physics~\cite{esslinger2016quantum,ritsch2013cold}. For instance they exhibit phase transitions among quantum phases such as Mott insulator (MI), superfluid (SF) and charge density wave~(CDW), which is a subject of foundational importance~\cite{bloch2002quantum,esslinger2016quantum,kovrizhin2005density,shimizu2018dynamics}. Phase transitions can be witnessed by investigating, e.g., energy gaps~\cite{bloch2002quantum,QPTBook2001}, short-range or long-range order parameters~\cite{esslinger2016quantum}. Another way to look into quantum phases is to examine the entanglement of ensembles, e.g. the behavior of entanglement across phase transitions~\cite{tasgin2017many,lambert2004entanglement,gunay2019entanglement}. Interest in entanglement features is not confined to the intriguing physics these collectively-entangled systems represent. Such controllable systems can provide a promising tool for on-demand generation of entanglement~\cite{gunay2023demand,enomoto2021programmable,ullah2023electrically}, which can be useful broadly in quantum information processing~\cite{optical_lattice_QC}. Entanglement induced across phase transitions can be stronger than those obtained e.g. via squeezing--type interactions \footnote{ For instance, in a superradiant or superfluid phase, the quasiparticle excitations exhibit number-squeezed correlation while in a spin-squeezed ensemble they show quadrature squeezed correlation~\cite{tasgin2017many}. Photon number states are known to possess nonclassicality stronger than the quadrature squeezed states. }.

Recent studies explored the entanglement charateristics substantially in different quantum phases and across phase transitions in optical lattices~\cite{frerot2016entanglement,walsh2019local,sharma2022quantum,zhang2016cluster,canella2019superfluid}. For instance, the approaches adopting slave-boson technique in ~\cite{frerot2016entanglement,walsh2019local,sharma2022quantum} show that the Mott insulator(MI)--superfluid (SF) phase transition is accompanied by a sharp singularity in the entanglement entropy for Bose-Hubbard model (BHM) manifesting quantum criticality~\cite{frerot2016entanglement,walsh2019local}. Those studies have also been extended to the case of optical lattice placed inside an optical cavity~\cite{sharma2022quantum}, which introduces long-range order leading to an extended BHM and thereby enriching the many-body dynamics, e.g. the emergence of CDW phase~\cite{sharma2022quantum,esslinger2016quantum,ritsch2013cold,ritsch2008ultracold}. 
These prior studies provide useful insights to understand quantum entanglement in strongly interacting quantum systems. However the analysis using entanglement entropy is not readily accessible experimentally unless a full quantum-state tomography is performed \footnote{D. F. V. James {\it et al.}, \pra {\bf 64}, 052312 (2001).}, which can be increasingly demanding with quantum many-body systems. There were also alternative approaches, e.g. measuring the Renyi-2 entropy, which provides only the lower-bound for entanglement entropy, based on the interference of two quantum copies~\footnote{R. Islam {\it et al.}, Nature {\bf528}, 77 (2015).} or the probability-correlations via random measurements~\footnote{T. Brydges {\it et al.} Science {\bf 364}, 260 (2019).}. Moreover, entanglement entropy characterizes the bipartite entanglement of the whole system not directly addressing collective nature of entanglement.

In this respect it is desired to establish an entanglement criterion that can simultaneously fulfill two conditions: (i) manifest and characterize collective entanglement adequately and (ii) be readily testable in practical settings. In this Letter, we introduce such a criterion for the characterization of collective entanglement generally applicable to Boson lattice systems. Our approach, which is based on an observable reflecting quantum interference among lattice sites, is further extended to come up with a hierarchy of multipartite entanglement conditions. We analyze collective entanglement due to the short-range~\cite{bloch2002quantum} 
 and long-range interactions in optical lattice ~\cite{esslinger2016quantum} to gain insight into quantum correlations revealed under our approach. We also show how our approach captures the aspects of phase transitions like MI-CDW and MI-SF crossovers in the extended BHM~\cite{PS_crossover}. 
Furthermore, we illustrate how our approach detects a hierarchy of multipartite entanglement present in lattice system.
 Our method can be experimentally accessed using existing techniques such as Raman scattering and Time-of-light~\cite{dao2007measuring,blakie2006raman,konabe2006laser,dao2009probing,leskinen2010resonant,hall1998dynamics,bernier2010thermometry,kozlowski2017quantum,mekhov2007light,kozlowski2015probing,javanainen1995off,gerbier2008expansion}. It therefore provides a powerful tool for the study of collective entanglement theoretically and practically in Bosonic systems.

{\small \bf Extended Bose-Hubbard model}--- We first introduce the model of the Boson lattice system. Dynamics of cold atoms in optical lattice may be addressed by the BHM ~\cite{ritsch2008ultracold}
\begin{equation}
	\hat{\cal H}_{\rm \scriptscriptstyle BH} = -J\sum_{\langle i, j\rangle} (\hat{b}_i^\dagger \hat{b}_j+H.c.) + \frac{U}{2} \sum_{i=1}^L  (\hat{b}_i^\dagger)^2 \hat{b}_i^2 ,
	\label{BHhamiltonian}
\end{equation} 
where $\hat{b}_i^\dagger$~($\hat{b}_i$) creates~(annihilates) a boson in the $i $th lattice site. $i,j=1,\cdots,L$ run through all $L$ number of lattice sites with $N$ a total number of bosons (trapped atoms) in the lattice. The first term accounts for the nearest-neighbor hopping between two sites $i$ and $j$ and the second term the on-site atom-atom interaction. Those two terms compete with each other to favor either the SF phase~with lattice sites entangled~\cite{walsh2019local} or MI phase with no entanglement, respectively, if the former interaction prevails over the latter or the opposite.

When the lattice system is placed in an optical cavity, a long-range interaction can be induced via coupling to a cavity field~\cite{ritsch2008ultracold} creating an additional Hamiltonian as
\begin{eqnarray}
	\hat{\cal H}_{\rm LR} = -\frac{U_{\rm \scriptscriptstyle LR}}{N} \left[ \sum_{i=1}^L (-1)^i \: \hat{n}_i \right]^2.
\end{eqnarray}  
That is, the atoms in the even and the odd sites interact collectively ($\hat{n}_i=\hat{b}_i^\dagger \hat{b}_i$:  occupation at site-$i$). 
$\hat{\cal H}_{\rm LR}$ can produce a charge density wave (CDW) phase observed in experiment~\cite{esslinger2016quantum}, which may be characterized by the order parameter $ \Theta_{\rm \scriptscriptstyle LR}=2\langle \sum_i (-1)^i \: \hat{n}_i  \rangle/N $ representing the disparity in population between even and odd sites~\cite{esslinger2016quantum}. 

 The state of lattice system realized in experiment may be well approximated as pure. This is because thermal fluctuations at $k_B T\sim 20$ Hz in such systems are extremely small compared to the relative energy scales, i.e., recoil or mean-interaction energies ~\cite{bloch2002quantum,esslinger2016quantum}, making the Boltzmann factors vanishingly small as $\exp(-{\varepsilon_i/k_BT})\propto \exp({-100})$.

{\small \bf  Entanglement criterion}--- We now present how the multipartite entanglement among boson lattice modes can be identified using an experimentally-accessible criterion based on a collective observable $\hat{\cal R}_{\bf q}$~\cite{sorensen2001Nature,DGCZ_PRL2000}. 
 Our guiding principle is to use quantum interference among boson lattice modes that can be reflected in the form of $\hat{\cal R}_{\bf q}$. For a separable state, its fluctuation is bounded by a value ${\cal R}_{\rm sep}$. A lower variance satisfying $\Delta^2\hat{\cal R}_{\bf q}  < {\cal R}_{\rm sep}$ thus confirms the many-body entanglement among lattice sites. 
We demonstrate the power of our approach by applying it to the extended-BHM, $\hat{\cal H}_{\rm \scriptscriptstyle BH} +\hat{\cal H}_{\rm LR}$.


We define a collective operator $\hat{B}_{\bf q}\equiv\frac{1}{\sqrt{L}} \sum_{j=1}^L \: e^{-\iota q_j} \: \hat{b}_j$~\cite{roy2022genuine}, where a phase $q_j$ at each site $j$ of mode $\hat{b}_j$ is introduced to obtain a wide range of entanglement criteria. That is, each set of ${\bf q}\equiv\{q_j\}$ defines a different observable making our approach flexible. Physically ${\bf q}$ may represent a quasi-momentum of the lattice system via the Fourier transform of lattice density distribution. 


Using the above $\hat{B}_{\bf q}$, we first try to construct a many-site entanglement criterion via a quadrature-like observable $\hat{B}_\phi\equiv(e^{\iota \phi}\hat{B}_{\bf q}+e^{-\iota\phi}\hat{B}_{\bf q}^\dagger)/\sqrt{2}$, where $\phi$ may run over $\left[0,2\pi\right]$. However it turns out not to witness the entanglement under the Mott insulator--superfluid and Mott insulator--CDW crossovers as shown in Supplementary Material (SM)~\cite{supplementary}. This is somewhat similar to a spin-squeezing criterion $\langle \hat{S}_x(\phi)\rangle$ \cite{duan2011many_particle_entanglement}, a Fermion analog of our Boson criterion $\hat{B}_\phi$, which does not witness the collective entanglement present in a superradiant phase~\cite{tasgin2017many}. 

We therefore consider a number-like observable, i.e. $\hat{\cal R}_{\bf q}=\hat{B}_{\bf q}^\dagger \hat{B}_{\bf q} $. 
With details in SM~\cite{supplementary}, we find that a separable state 
	$\hat{\rho}_{\rm sep}=\sum_k P_k \: \rho_1^{(k)} \otimes \rho_2^{(k)} \otimes \ldots \otimes \rho_L^{(k)} ,
	\label{rho_sep}$
has a lower-bound ${\cal R}_{\rm sep}$ for its variance of $\hat{\cal R}_{\bf q}$, that is $\Delta^2\hat{\cal R}_{\bf q} \ge {\cal R}_{\rm sep}$, given by
\begin{equation}
	{\cal R}_{\rm  sep} = \Big[  N(L-1) +N^2 -\sum_{j=1}^{L} \langle \hat{n}_j \rangle^2 \Big] / L^2.
	\label{Rsep}
\end{equation}
Here $N$ is the total number of atoms, $L$ the number of lattice sites, with the last term $\langle \hat{n}_j \rangle$ accessible via occupation number imaging experiments~\cite{bakr2009quantum,sherson2010single,campbell2006imaging,gerbier2006probing}.
Note that the separability bound ${\cal R}_{\rm  sep}$ is independent of the choice ${\bf q}=\{q_j\}$ in the observable $\hat{\cal R}_{\bf q}$~\cite{supplementary}.   
Therefore, if the condition 
\begin{equation}
	r\equiv\frac{\Delta^2\hat{\cal R}_{\bf q}}{{\cal R}_{\rm  sep}}<1,
    \label{EC}
\end{equation}
is satisfied for any choice of phase setting ${\bf q}=\{q_j\}$, it verifies quantum entanglement among the lattice sites. As we show by examples, a given entangled state satisfies the entanglement condition~(\ref{EC}) for a broad range of ${\bf q}=\{q_j\}$ thereby enabling entanglement detection experimentally favorable.




{\small \bf  Entanglement in extended BHM}---We illustrate our criterion by investigating the collective entanglement in the ground states of extended BHM ~\cite{frerot2016entanglement,walsh2019local,sharma2022quantum}. 
First, to gain insight into how our approach characterizes entanglement for different quantum phases, we analytically study two prototype phases, i.e. SF and CDW phases. The former arises prominently when there is only the short-range interaction (hopping) without the long-range interaction, whereas the latter does in the opposite limit. By studying these two phases, we can thus grasp the characteristics of entanglement owing to the short-range and the long-range interactions, respectively, which also provides the basis to understand the cases with both interactions present.

With details in SM ~\cite{supplementary}, SF phase manifests a stronger multipartite entanglement than CDW phase. For instance, a SF state $|\psi_{\rm SF} \rangle = \frac{1}{\sqrt{N!}}\bigg( \frac{1}{\sqrt{L}}\sum_{j=1}^L \hat{b}_j^\dagger \bigg)^N |0\rangle$ having an equal population among lattice sites with coherence gives a completely zero variance $\Delta^2\hat{\cal R}_{\bf q=0}=0$ at the choice of phase setting ${\bf q}=\{q_j=0\}$ in the observable $\hat{\cal R}_{\bf q}$. In addition, the variance $\Delta^2\hat{\cal R}_{\bf q}$ is negligibly small in the vicinity of the choice ${\bf q}=\{q_j=0\}$, therefore making its experimental verification favorable. In Fig. 1, we show how our criterion manifests the entanglement of the SF phase in 2D optical lattice. As discussed before, each phase setting ${\bf q}=\{q_j\}$ defines a distinct observable $\hat{\cal R}_{\bf q}$. In this 2D case, each point $\{q_x,q_y\}$ in the parameter space represents a different observable and verifies entanglement if the condition $r=\frac{\Delta^2 \hat{R}_q}{R_{\rm sep}}<1$ is met. In Fig. 1 (a), we show the intensity pattern of the signal $\langle \hat{R}_{\bf q}\rangle$ as a function of $q_x$ and $q_y$. More importantly, in Fig. 1 (b), we present a contour plot for the value $r=\frac{\Delta^2 \hat{R}_q}{R_{\rm sep}}$ representing the variance of the signal at each $\{q_x,q_y\}$. We see that not only at $\{q_x=0,q_y=0\}$ but also in a broad range of $\{q_x,q_y\}$, the condition $r=\frac{\Delta^2 \hat{R}_q}{R_{\rm sep}}<1$ is satisfied thereby manifesting entanglement successfully. As addressed later, each $\{q_x,q_y\}$ may experimentally refer to either the wave vector of the signal in Raman scattering or the momentum in time of flight (Fig. 4). Therefore, entanglement is verified in a broad range of experimental conditions enabling a practical test favorable. 
In SM~\cite{supplementary}, we also discuss a statistical error in real experiment and demonstrate that a successful detection of entanglement is still possible even adding that effect.

On the other hand, the CDW phase represents at most 1-ebit (entangled bit) entanglement between even and odd sites collectively~\cite{supplementary}, which would become relatively negligible with respect to the scale of lattice size $L$ or atom number $N$. Another noteworthy point is that CDW phase gives a variance {\it completely insensitive} to the choice of ${\bf q}=\{q_j\}$ in the observable $\hat{\cal R}_{\bf q}$. That is, its detection is successful with the same value $r\equiv\frac{\Delta^2\hat{\cal R}_{\bf q}}{{\cal R}_{\rm  sep}}$ regardless of ${\bf q}=\{q_j\}$~\cite{supplementary}. 

The above results for SF and CDW phases combined together may give us some clues to understand the cases with both short-range and long-range interactions present. For instance, we show in SM~\cite{supplementary} that the phase-sensitivity for entanglement detection diminishes with increasing $U_{\rm LR}$ for a fixed $J/U$. We also show later in main text how the detection of multipartite entanglement can be influenced by both interactions.

\begin{figure}
	\begin{center}
	       \includegraphics[width=0.485\textwidth]{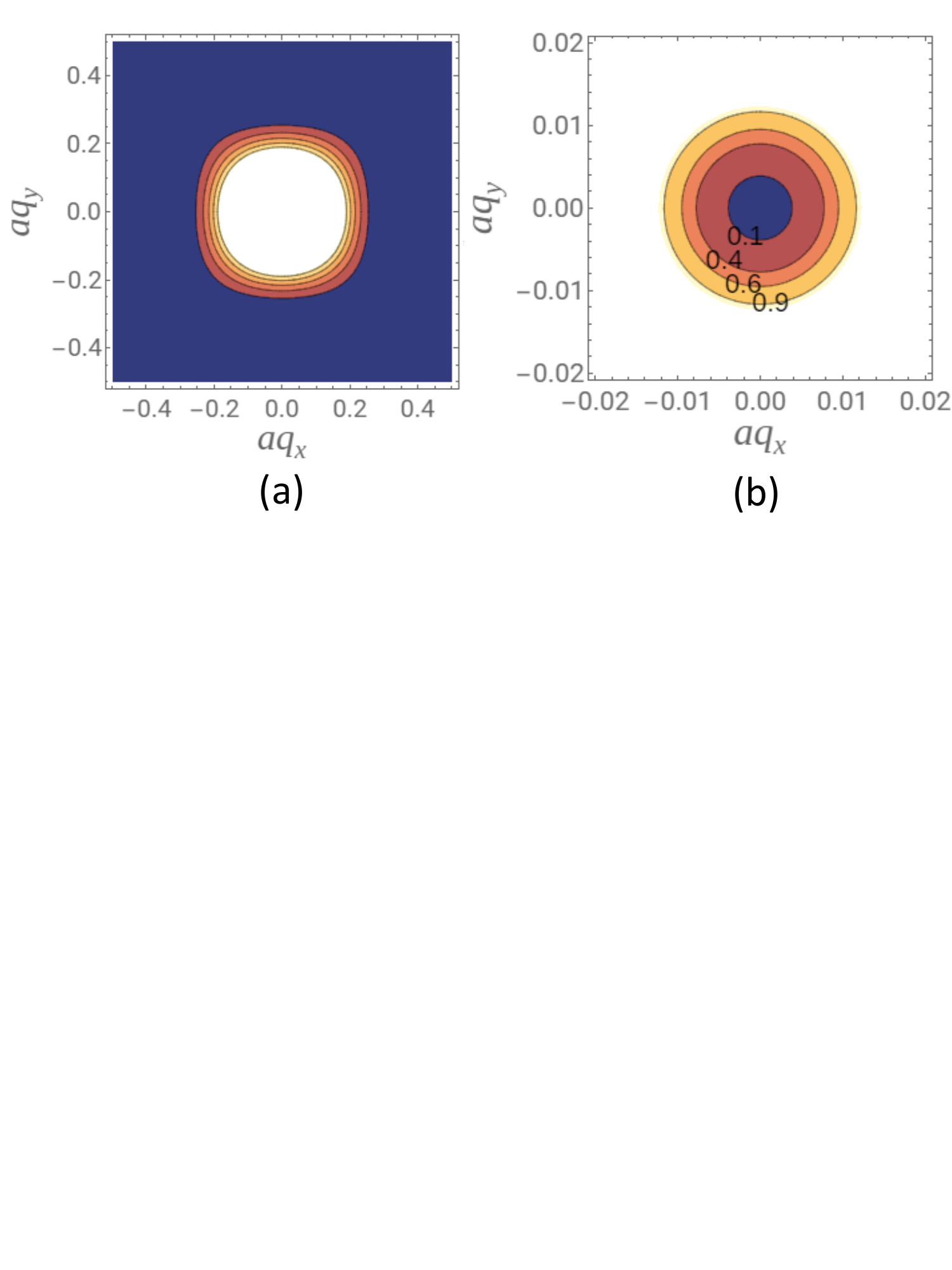}
    \vspace{-7.5cm}
		\caption{Observing entanglement for SF phase in 2D lattice of size $L=20\times20$ with atom number $N=L$ and lattice constant $a$ via our criterion in Eq. (4). Dark to bright color indicates the increase of values. (a) Intensity pattern of the signal $\langle \hat{R}_{\bf q}\rangle$ as a function of $q_xa$ and $q_ya$, where $(q_x,q_y)$ may experimentally refer to either the wave vector of the signal in Raman scattering or the momentum in time of flight (Fig. 4). (b) Contour plot representing the value of ratio $\Delta^2 \hat{R}_{\bf q}/R_{\rm sep}$, whose value below 1 is the signature of entanglement. We see that entanglement is verified in a broad range of parameters $\{q_x,q_y\}$ for the observable $\hat{R}_{\bf q}$. }
		\label{fig4}
	\end{center}
\end{figure}

\begin{figure}
	\begin{center}
		\includegraphics[width=0.45\textwidth]{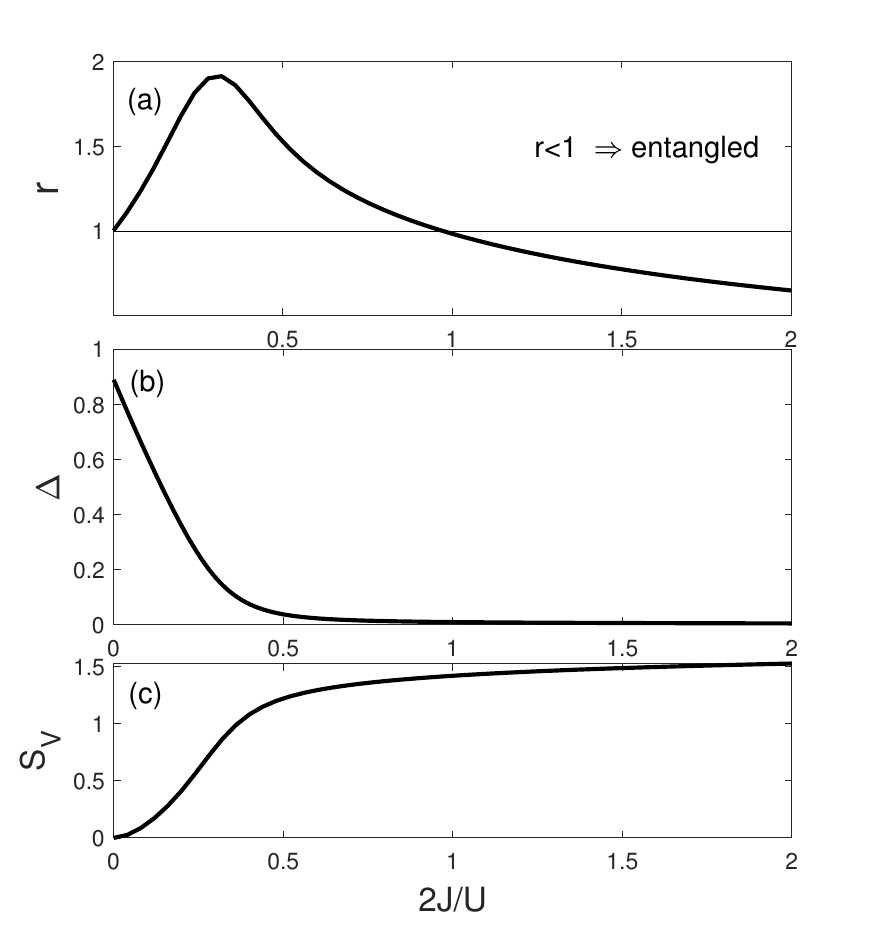}
		\caption{Onset of entanglement among lattice sites for MI to SF phase transition. (a) Our entanglement witness $r\equiv\frac{\Delta^2\hat{\cal R}_{\bf q}}{{\cal R}_{\rm  sep}}$ with a choice $\{q_j=0\}$ falls down below 1 at about $2J=U$ where (b) energy gap $\Delta$ becomes almost zero indicating a crossover to the SF phase~\cite{bloch2002quantum,QPTBook2001}. (c) Entanglement entropy between the first $N/2$ and last $N/2$ sites increasing monotonically with the hopping parameter $J$. In this case, $S_V$ does not capture the phase transition unlike our criterion. Exact calculations have been performed for $L=N=8$.} 
		\label{fig2}
	\end{center}
\end{figure}

{\it Phase Transitions}---We next show how our criterion characterizes the phase transitions in the extended BHM. 
In Fig.~\ref{fig2}, we study the MI-SF crossover in the absence of long range interaction $U_{\rm \scriptscriptstyle LR}=0$. 
For this investigation we also consider an energy gap defined by 
$\Delta(N)=N\left[ E(N+1)/(N+1) + E(N-1)/(N-1) -2\: E(N)/N \right]$ \cite{kelecs2015mott,cooper2001quantum}, where $E(N+1)$, $E(N)$ and $E(N-1)$ are the ground state energies evaluated with $N$ number of bosons. A Mott insulator state is known to be characterized by a finite gap while the gap vanishes for the superfluid phase~\cite{bloch2002quantum,QPTBook2001}, thus making $\Delta$ a useful tool to examine phase transition.

We observe that energy gap $\Delta$ approaches quite close to zero (Fig.~\ref{fig2}b) when $r\equiv\frac{\Delta^2\hat{\cal R}_{\bf q}}{{\cal R}_{\rm  sep}}$ falls down below 1 at about $2J=U$ (Fig.~\ref{fig2}a).  We mention that insulator-to-superfluid crossover takes place smoothly in our simulation with a small number of bosons ($N=8$) while it becomes an abrupt transition for a large $N$. 
In comparison, from Fig.~\ref{fig2}c, the entanglement entropy between the first $4$ sites and the last $4$ sites monotonically increases with $J$ without displaying the nature of phase transition unlike our approach. 

We also analyze the MI--CDW crossover~\cite{PS_crossover} by letting the hopping interaction $J=0$ in Sec. V of SM~\cite{supplementary}. We show that our criterion captures well the transition between the two phases via the condition $r\equiv\frac{\Delta^2\hat{\cal R}_{\bf q}}{{\cal R}_{\rm  sep}}<1$ after the transition at $U_{\rm \scriptscriptstyle LR} = 0.5 U$. The cases with the short-range and the long-range interactions present together are discussed below and in SM~\cite{supplementary}, particularly in relation to multipartite entanglement.

{\small \bf Multipartite entanglement}---
We now extend our approach to reveal a multipartite nature of entanglement among boson lattice modes. Although we demonstrated the usefulness of our entanglement criterion in different situations, Eq.~(\ref{EC}) does not certify multipartite entanglement rigorously because its derivation is based on the negation of a fully separable state $\hat{\rho}_{\rm sep}=\sum_k P_k \: \rho_1^{(k)} \otimes \rho_2^{(k)} \otimes \ldots \otimes \rho_L^{(k)}$. We may further consider a mixture of states with a partial entangled structure, e.g. $\hat{\rho}_{\rm PE}=\sum_k P_k \: \rho_{1,\cdots,L_1}^{(k)} \otimes \rho_{L_1+1,\cdots,L_2}^{(k)} \otimes \ldots \otimes \rho_{L_m+1,\cdots,L}^{(k)}$, and the bound arising from such a state to find a criterion on multipartite nature of entanglement.

Let us suppose that the maximum number of modes entangled with one another is at most $L_1$ out of total $L$-modes ($L_1>L/2$) in the decomposition of a given state. Then, instead of using a single observable $\hat{\cal R}_{\bf q}$, we consider $L_1$ number of observables, i.e. $\sum_{m=1}^{L_1} \Delta^2 \hat{\cal R}_{{\bf q}_m}$. As shown in SM~\cite{supplementary}, if $\hat{\cal R}_{{\bf q}_m}$'s are chosen such that the collective operators addressing the entangled $L_1$-modes are orthogonal to each other, e.g. Fourier modes, the sum of variances is lower-bounded as 
\begin{eqnarray}
\sum_{m=1}^{L_1} \Delta^2 \hat{\cal R}_{{\bf q}_m}\ge\frac{L_1(L-L_1)}{L^2}N\equiv {\cal R}^{L_1}. 
\label{ME}
\end{eqnarray}
That is, if Eq.~(\ref{ME})
is violated, it manifests $L_1+1$-partite entanglement. For instance, if we want to reveal the collective entanglement among all $L$-sites, we may perform a test with $L_1=L-1$ in Eq.~(\ref{ME}).

In Fig. 3, we illustrate how our test reveals multipartite entanglement in the extended BHM. For each test, we calculate the ratio $R_{\rm sum}=\frac{\sum_{m=1}^{L_1} \Delta^2 \hat{\cal R}_{{\bf q}_m}}{{\cal R}^{L_1}}$ whose value below 1 manifests $L_1+1$-partite entanglement. As we see from Fig. 3 (a) and (b), the level of multipartite entanglement becomes deeper with the 
 nearest-neighbor interaction $2J/U$ increasing for a given $U_{\rm LR}$. The long-range interaction $U_{\rm LR}$ may have a cooperating effect with the hopping $J$ when $U_{\rm LR}$ is rather small. In Fig. 3 (c), we show the case of testing a full $L=8$-partite entanglement, which manifests at a slightly less $2J/U$ and then at a higher $2J/U$ as $U_{\rm LR}$ increases. As discussed before, this may be attributed to the characteristic of entanglement in CDW phase, which arises due to the unequal population between even-odd sites collectively, creating at most 1-ebit entanglement regardless of lattice size or atom numbers.

\begin{figure}
	\begin{center}
		\includegraphics[width=0.56\textwidth]{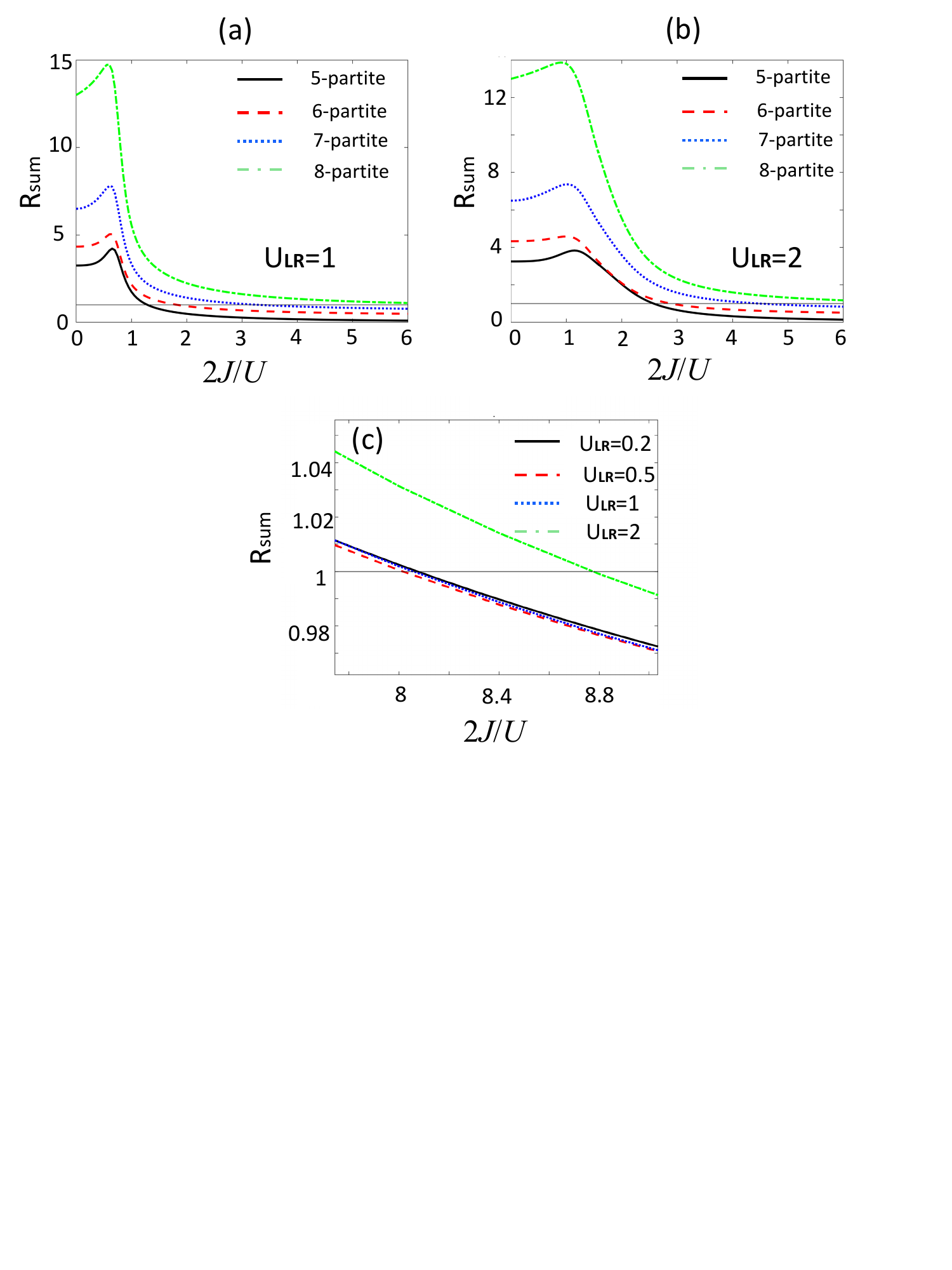}
        \vspace{-5.8cm}
         
		\caption{Certifying multipartite entanglement using $R_{\rm sum}=\frac{\sum_{m=1}^{L_1} \Delta^2 \hat{\cal R}_{{\bf q}_m}}{{\cal R}^{L_1}}$, whose value below 1 manifests $L_1+1$ partite entanglement, as a function of $2J/U$ for (a) $U_{\rm LR}=1$ and (b) $U_{\rm LR}=2$. Plot (c) shows the results for testing a full $L$-partite entanglement with varied $U_{\rm LR}$. }
		\label{fig4}
	\end{center}
\end{figure}

{\small \bf Measuring $\hat{\cal R}_{\bf q}$ in experiments}--- Not only capturing the characteristics of collective entanglement in Boson lattice systems, our criterion can also be readily tested using established experimental techniques. 
The observable $\hat{\cal R}_{\bf q}$ can be accessed specifically via Raman-scattering ~\cite{dao2007measuring,kozuma1999coherent,blakie2006raman,konabe2006laser,dao2009probing} and time-of-flight method~\cite{gerbier2008expansion} facilitating quantum interference among lattice sites~\cite{polkovnikov2006interference,niu2006imaging}.  
 
{\it Raman scattering}---In the Raman process~\cite{dao2007measuring,kozuma1999coherent,jimenez2012peierls}, an initial lattice state~(trapped state) can be recoiled into another internal state~(referred to as free state)~\cite{hall1998dynamics}. This enables the observation of one-particle correlation $C({\bf r},{\bf {r}'})=\langle \hat{\psi}^\dagger({\bf r},t) \hat{\psi}({\bf r'},t')\rangle$~\cite{dao2007measuring,kozuma1999coherent,schafer2020tools} where $\hat{\psi}({\bf r},t)$ is the trapped state operator. Raman intensity $I_{\rm \scriptscriptstyle Ram}$ is proportional to $I_{\rm \scriptscriptstyle Ram} \propto \int d^3{\bf r} \int d^3{\bf r}' e^{ -\iota {\bf q}\cdot ({\bf r}-{\bf r}') }   \psi_2({\bf r}) \psi_2^*({\bf r }') \langle \hat{\psi}^\dagger({\bf r},t) \hat{\psi}({\bf r'},t')\rangle $~\cite{blakie2006raman,konabe2006laser,dao2009probing} where $\psi_2({\bf r})$ is the wave-function corresponding to the second internal state.  When the trapping for the free state is sufficiently larger than the inter-species interaction  $U_{12}$, which can also be tuned by Feshbach resonances~\cite{inouye1998observation,theis2004tuning}, $\psi_2({\bf r})$ are Bloch-Flouqet solutions. Thus, employing the usual Wannier expansion $\hat{\psi}({\bf r})=\sum_j w_j ({\bf r})$ $\hat{b}_j$~\cite{ritsch2008ultracold,blakie2006raman}, we obtain $\hat{\cal R}_{\bf q}=\sum_i \sum_j e^{\iota \vec{q}\cdot ({\bf r}_i-{\bf r}_j)} \hat{b}_i^\dagger \hat{b}_j$ from $I_{\rm \scriptscriptstyle Ram}$. Here $\vec{q}=\vec{k}_1-\vec{k}_2$ corresponds to the difference in wave vector between two Raman fields as shown in Fig. 4 (a), and is related to the phase $q_j=\vec{q}\cdot\vec{r}_j$ in our collective operator $\hat{B}_{\bf q}\equiv\frac{1}{\sqrt{L}} \sum_{j=1}^L \: e^{-\iota q_j} \: \hat{b}_j$ with $\hat{\cal R}_{\bf q}=\hat{B}^\dag_{\bf q}\hat{B}_{\bf q}$. 

\begin{figure}
	\begin{center}
		\includegraphics[width=0.5\textwidth]{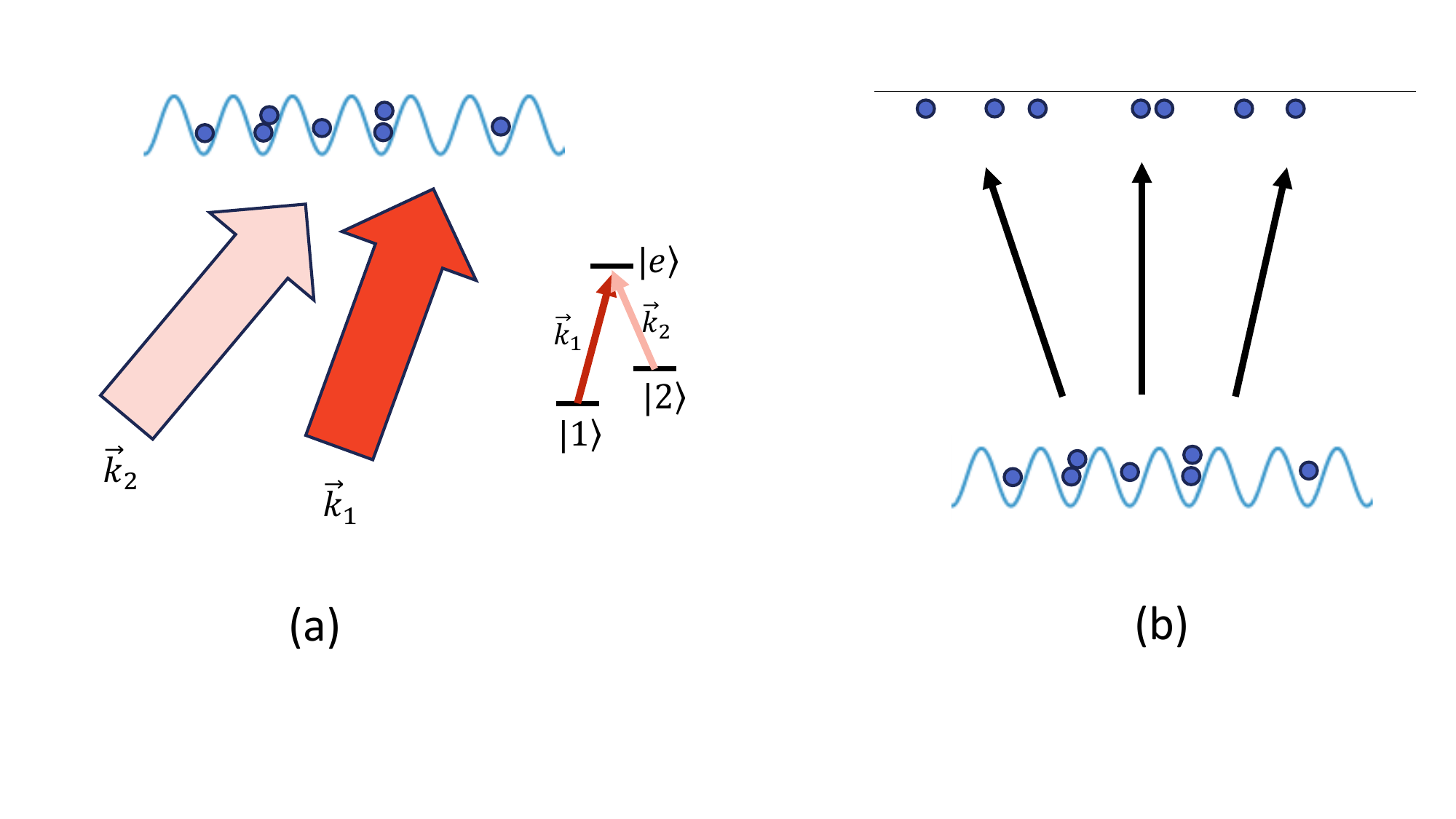}
        \vspace{-1.5cm}
         
		\caption{Two experimental schemes to observe $\hat{\cal R}_{\bf q}$ (a) Raman scattering with the momentum $\vec{q}=\vec{k}_1-\vec{k}_2$ of our observable $\hat{\cal R}_{\bf q}$ determined by the wave vectors of two Raman fields  (b) Time of flight observing the position distribution of atoms at a far distance that essentially measures the momentum $\vec{q}$ of the initial lattice state. }
		\label{fig4}
	\end{center}
\end{figure}

We note that in the Raman scattering, coefficients of  $\hat{b}_i^\dagger \hat{b}_j$ terms are equally distributed apart from a phase factor. On the other hand, this is not the case with the Bragg scattering~\cite{miyake2011bragg} where recoil takes place into the same trapped state, thus probing two-particle correlation $\langle \hat{\psi}^\dagger({\bf r},t)  \hat{\psi}({\bf r},t)  \hat{\psi}^\dagger({\bf r}',t') \hat{\psi}({\bf r'},t')\rangle$~\cite{dao2007measuring,javanainen1995off}. In this case, to access the observable $\hat{\cal R}_{\bf q}$, one needs to investigate electric field $\hat{E}_{\rm \scriptscriptstyle sca}=\hat{E}^{(+)} + \hat{E}^{(-)}$ instead of the intensity $I_{\rm \scriptscriptstyle sca}= \langle  \hat{E}^{(+)}  \hat{E}^{(-)} \rangle$. Positive frequency part of the scattered field is $\hat{E}^{(+)} \propto \int d^3 {\bf r} e^{-\iota {\bf q} \cdot {\bf r}} \hat{\psi}^\dagger ({\bf r}) \hat{\psi} ({\bf r})   $~\cite{javanainen1995off} where $ {\bf q}$ is the change in momentum after the scattering.
In this $\hat{E}^{(+)}$, however, arguments of  $\hat{\psi}^\dagger ({\bf r})$ and $\hat{\psi} ({\bf r}')$ operators are the same, ${\bf r}'={\bf r}$, unlike the Raman intensity $I_{\rm \scriptscriptstyle Ram}$. When we carry out the Wannier expansion and the ${\bf r}$ integration, $\hat{E}^{(+)} \propto (\hat{D}+\hat{B}) $~\cite{kozlowski2017quantum,mekhov2007light,kozlowski2015probing} images only the onsite $\hat{D}=\sum_{j=1}^L \hat{b}_j^\dagger \hat{b}_j$ and the nearest-neighbor $\hat{B}=\sum_{j=1}^{M-1} (\hat{b}_{j+1}^\dagger \hat{b}_j + H.c. ) $ correlations. We could obtain an entanglement criterion also by investigating e.g. the variance of $\hat{P}=\hat{D}+\hat{B}$ operator, 
which would describe the lattice correlations in a different context from $\hat{\cal R}_{\bf q}$. 

%

{\it Time of flight}---$\hat{\cal R}_{\bf q}$ can also be accessed via time-of-flight measurement~\cite{gerbier2008expansion}. The atomic density $n_{\rm ToF}(\vec{r})$  observed after the free expansion of the trapped atoms from optical lattice is determined by the spatial coherence of lattice sites, i.e. $S({\bf q})=\sum_{i,j} e^{i {\bf q}\cdot ({\bf r}_i-\bf{r}_j)}  \langle \hat{b}_i^\dagger \hat{b}_j \rangle $ \cite{gerbier2008expansion}. Letting $q_i\equiv {\bf q}\cdot {\bf r}_i$, we see that the interference term $S({\bf q})$ is  the observable $\hat{\cal R}_{\bf q}$ to measure in our criterion, $S({\bf q})\propto \hat{\cal R}_{\bf q}=\hat{B}^\dag_{\bf q}\hat{B}_{\bf q}$ (Fig. 4 (b)).  For instance, the intensity profile shown in Fig. 1 (a) may be obtained by the atomic density distribution at a far field in time-of-flight method. Testing our criterion then corresponds to analyzing the fluctuation of the signal at each point.  

In conclusion, we have presented an observable entanglement criterion that can manifest the collective entanglement in Boson lattice systems. Our criterion captures the characteristics of entanglement in different quantum phases and across phase transitions well in the extended Bose-Hubbard system, thereby enabling a comprehensive study of many-body physics in the context of quantum correlations. Furthermore, it can be readily tested using available experimental techniques providing a useful tool of fundamental and practical importance that we hope will enrich future studies of quantum many-body physics. For instance, our tool can be utilized to verify on-demand generation of entanglement for programmable quantum information processing. Once the nature of the entanglement is identified, one can use it on its own or further design an efficient interaction Hamiltonian to transfer lattice entanglement to other systems. 

For decades, the spin-squeezing criterion has been extensively employed in the study of {\it multi-particle} entanglement in spin-systems (two-level atoms) not only for foundational aspects ~\cite{sorensen2001Nature, Review, Mol}\footnote{D. J. Wineland, J. J. Bollinger, W. M. Itano, F. L. Moore, and
D. J. Heinzen, Spin squeezing and reduced quantum noise in
spectroscopy, Phys. Rev. A {\bf 46}, R6797 (1992).} but also for quantum applications, e.g. relation of multipartite entanglement to quantum metrological power~\footnote{P. Hyllus, L. Pezzé, and A. Smerzi, Entanglement and Sensitivity
in Precision Measurements with States of a Fluctuating
Number of Particles, Phys. Rev. Lett. {\bf 105}, 120501 (2010).
}\cite{Tot}. Likewise, we hope our proposed criterion could be a valuable tool for the study of {\it multi-mode} collective entanglement broadly in Bosonic quantum systems paving a new avenue for future studies. 

{\bf \small Acknowledgments.}--- We thank Mehmet \"Ozg\"ur Oktel, Ahmet Kele\c{s} and Sururi Emre Ta\c{s}c{\i} for their help with numerical calculations and discussions. MET acknowledges support from T\"UBA GEBIP 2017 and TUBITAK-1001 grant no 121F141. HN is partly supported by NRF-2022M3H3A1098237 from National Research Foundation of Korea.


\bibliography{bibliography}

\begin{thebibliography}{61}%
\makeatletter
\providecommand \@ifxundefined [1]{%
 \@ifx{#1\undefined}
}%
\providecommand \@ifnum [1]{%
 \ifnum #1\expandafter \@firstoftwo
 \else \expandafter \@secondoftwo
 \fi
}%
\providecommand \@ifx [1]{%
 \ifx #1\expandafter \@firstoftwo
 \else \expandafter \@secondoftwo
 \fi
}%
\providecommand \natexlab [1]{#1}%
\providecommand \enquote  [1]{``#1''}%
\providecommand \bibnamefont  [1]{#1}%
\providecommand \bibfnamefont [1]{#1}%
\providecommand \citenamefont [1]{#1}%
\providecommand \href@noop [0]{\@secondoftwo}%
\providecommand \href [0]{\begingroup \@sanitize@url \@href}%
\providecommand \@href[1]{\@@startlink{#1}\@@href}%
\providecommand \@@href[1]{\endgroup#1\@@endlink}%
\providecommand \@sanitize@url [0]{\catcode `\\12\catcode `\$12\catcode
  `\&12\catcode `\#12\catcode `\^12\catcode `\_12\catcode `\%12\relax}%
\providecommand \@@startlink[1]{}%
\providecommand \@@endlink[0]{}%
\providecommand \url  [0]{\begingroup\@sanitize@url \@url }%
\providecommand \@url [1]{\endgroup\@href {#1}{\urlprefix }}%
\providecommand \urlprefix  [0]{URL }%
\providecommand \Eprint [0]{\href }%
\providecommand \doibase [0]{https://doi.org/}%
\providecommand \selectlanguage [0]{\@gobble}%
\providecommand \bibinfo  [0]{\@secondoftwo}%
\providecommand \bibfield  [0]{\@secondoftwo}%
\providecommand \translation [1]{[#1]}%
\providecommand \BibitemOpen [0]{}%
\providecommand \bibitemStop [0]{}%
\providecommand \bibitemNoStop [0]{.\EOS\space}%
\providecommand \EOS [0]{\spacefactor3000\relax}%
\providecommand \BibitemShut  [1]{\csname bibitem#1\endcsname}%
\let\auto@bib@innerbib\@empty
\bibitem [{Note1()}]{Note1}%
  \BibitemOpen
  \bibinfo {note} {C. Gross and W. S. Bakr, Quantum gas microscopy for single
  atom and spin detection, Nature Phys. {\protect \bf 17}, 1316
  (2021).}\BibitemShut {Stop}%
\bibitem [{\citenamefont {Inouye}\ \emph {et~al.}(1998)\citenamefont {Inouye},
  \citenamefont {Andrews}, \citenamefont {Stenger}, \citenamefont {Miesner},
  \citenamefont {Stamper-Kurn},\ and\ \citenamefont
  {Ketterle}}]{inouye1998observation}%
  \BibitemOpen
  \bibfield  {author} {\bibinfo {author} {\bibfnamefont {S.}~\bibnamefont
  {Inouye}}, \bibinfo {author} {\bibfnamefont {M.}~\bibnamefont {Andrews}},
  \bibinfo {author} {\bibfnamefont {J.}~\bibnamefont {Stenger}}, \bibinfo
  {author} {\bibfnamefont {H.-J.}\ \bibnamefont {Miesner}}, \bibinfo {author}
  {\bibfnamefont {D.~M.}\ \bibnamefont {Stamper-Kurn}},\ and\ \bibinfo {author}
  {\bibfnamefont {W.}~\bibnamefont {Ketterle}},\ }\bibfield  {title} {\bibinfo
  {title} {Observation of feshbach resonances in a bose--einstein condensate},\
  }\href@noop {} {\bibfield  {journal} {\bibinfo  {journal} {Nature}\ }\textbf
  {\bibinfo {volume} {392}},\ \bibinfo {pages} {151} (\bibinfo {year}
  {1998})}\BibitemShut {NoStop}%
\bibitem [{\citenamefont {Theis}\ \emph {et~al.}(2004)\citenamefont {Theis},
  \citenamefont {Thalhammer}, \citenamefont {Winkler}, \citenamefont {Hellwig},
  \citenamefont {Ruff}, \citenamefont {Grimm},\ and\ \citenamefont
  {Denschlag}}]{theis2004tuning}%
  \BibitemOpen
  \bibfield  {author} {\bibinfo {author} {\bibfnamefont {M.}~\bibnamefont
  {Theis}}, \bibinfo {author} {\bibfnamefont {G.}~\bibnamefont {Thalhammer}},
  \bibinfo {author} {\bibfnamefont {K.}~\bibnamefont {Winkler}}, \bibinfo
  {author} {\bibfnamefont {M.}~\bibnamefont {Hellwig}}, \bibinfo {author}
  {\bibfnamefont {G.}~\bibnamefont {Ruff}}, \bibinfo {author} {\bibfnamefont
  {R.}~\bibnamefont {Grimm}},\ and\ \bibinfo {author} {\bibfnamefont {J.~H.}\
  \bibnamefont {Denschlag}},\ }\bibfield  {title} {\bibinfo {title} {Tuning the
  scattering length with an optically induced feshbach resonance},\ }\href@noop
  {} {\bibfield  {journal} {\bibinfo  {journal} {Physical Review Letters}\
  }\textbf {\bibinfo {volume} {93}},\ \bibinfo {pages} {123001} (\bibinfo
  {year} {2004})}\BibitemShut {NoStop}%
\bibitem [{\citenamefont {Sch{\"a}fer}\ \emph {et~al.}(2020)\citenamefont
  {Sch{\"a}fer}, \citenamefont {Fukuhara}, \citenamefont {Sugawa},
  \citenamefont {Takasu},\ and\ \citenamefont {Takahashi}}]{schafer2020tools}%
  \BibitemOpen
  \bibfield  {author} {\bibinfo {author} {\bibfnamefont {F.}~\bibnamefont
  {Sch{\"a}fer}}, \bibinfo {author} {\bibfnamefont {T.}~\bibnamefont
  {Fukuhara}}, \bibinfo {author} {\bibfnamefont {S.}~\bibnamefont {Sugawa}},
  \bibinfo {author} {\bibfnamefont {Y.}~\bibnamefont {Takasu}},\ and\ \bibinfo
  {author} {\bibfnamefont {Y.}~\bibnamefont {Takahashi}},\ }\bibfield  {title}
  {\bibinfo {title} {Tools for quantum simulation with ultracold atoms in
  optical lattices},\ }\href@noop {} {\bibfield  {journal} {\bibinfo  {journal}
  {Nature Reviews Physics}\ }\textbf {\bibinfo {volume} {2}},\ \bibinfo {pages}
  {411} (\bibinfo {year} {2020})}\BibitemShut {NoStop}%
\bibitem [{\citenamefont {Landig}\ \emph {et~al.}(2016)\citenamefont {Landig},
  \citenamefont {Hruby}, \citenamefont {Dogra}, \citenamefont {Landini},
  \citenamefont {Mottl}, \citenamefont {Donner},\ and\ \citenamefont
  {Esslinger}}]{esslinger2016quantum}%
  \BibitemOpen
  \bibfield  {author} {\bibinfo {author} {\bibfnamefont {R.}~\bibnamefont
  {Landig}}, \bibinfo {author} {\bibfnamefont {L.}~\bibnamefont {Hruby}},
  \bibinfo {author} {\bibfnamefont {N.}~\bibnamefont {Dogra}}, \bibinfo
  {author} {\bibfnamefont {M.}~\bibnamefont {Landini}}, \bibinfo {author}
  {\bibfnamefont {R.}~\bibnamefont {Mottl}}, \bibinfo {author} {\bibfnamefont
  {T.}~\bibnamefont {Donner}},\ and\ \bibinfo {author} {\bibfnamefont
  {T.}~\bibnamefont {Esslinger}},\ }\bibfield  {title} {\bibinfo {title}
  {Quantum phases from competing short-and long-range interactions in an
  optical lattice},\ }\href@noop {} {\bibfield  {journal} {\bibinfo  {journal}
  {Nature}\ }\textbf {\bibinfo {volume} {532}},\ \bibinfo {pages} {476}
  (\bibinfo {year} {2016})}\BibitemShut {NoStop}%
\bibitem [{\citenamefont {Ritsch}\ \emph {et~al.}(2013)\citenamefont {Ritsch},
  \citenamefont {Domokos}, \citenamefont {Brennecke},\ and\ \citenamefont
  {Esslinger}}]{ritsch2013cold}%
  \BibitemOpen
  \bibfield  {author} {\bibinfo {author} {\bibfnamefont {H.}~\bibnamefont
  {Ritsch}}, \bibinfo {author} {\bibfnamefont {P.}~\bibnamefont {Domokos}},
  \bibinfo {author} {\bibfnamefont {F.}~\bibnamefont {Brennecke}},\ and\
  \bibinfo {author} {\bibfnamefont {T.}~\bibnamefont {Esslinger}},\ }\bibfield
  {title} {\bibinfo {title} {Cold atoms in cavity-generated dynamical optical
  potentials},\ }\href@noop {} {\bibfield  {journal} {\bibinfo  {journal}
  {Reviews of Modern Physics}\ }\textbf {\bibinfo {volume} {85}},\ \bibinfo
  {pages} {553} (\bibinfo {year} {2013})}\BibitemShut {NoStop}%
\bibitem [{\citenamefont {Greiner}\ \emph {et~al.}(2002)\citenamefont
  {Greiner}, \citenamefont {Mandel}, \citenamefont {Esslinger}, \citenamefont
  {H{\"a}nsch},\ and\ \citenamefont {Bloch}}]{bloch2002quantum}%
  \BibitemOpen
  \bibfield  {author} {\bibinfo {author} {\bibfnamefont {M.}~\bibnamefont
  {Greiner}}, \bibinfo {author} {\bibfnamefont {O.}~\bibnamefont {Mandel}},
  \bibinfo {author} {\bibfnamefont {T.}~\bibnamefont {Esslinger}}, \bibinfo
  {author} {\bibfnamefont {T.~W.}\ \bibnamefont {H{\"a}nsch}},\ and\ \bibinfo
  {author} {\bibfnamefont {I.}~\bibnamefont {Bloch}},\ }\bibfield  {title}
  {\bibinfo {title} {Quantum phase transition from a superfluid to a mott
  insulator in a gas of ultracold atoms},\ }\href@noop {} {\bibfield  {journal}
  {\bibinfo  {journal} {Nature}\ }\textbf {\bibinfo {volume} {415}},\ \bibinfo
  {pages} {39} (\bibinfo {year} {2002})}\BibitemShut {NoStop}%
\bibitem [{\citenamefont {Kovrizhin}\ \emph {et~al.}(2005)\citenamefont
  {Kovrizhin}, \citenamefont {Pai},\ and\ \citenamefont
  {Sinha}}]{kovrizhin2005density}%
  \BibitemOpen
  \bibfield  {author} {\bibinfo {author} {\bibfnamefont {D.~L.}\ \bibnamefont
  {Kovrizhin}}, \bibinfo {author} {\bibfnamefont {G.~V.}\ \bibnamefont {Pai}},\
  and\ \bibinfo {author} {\bibfnamefont {S.}~\bibnamefont {Sinha}},\ }\bibfield
   {title} {\bibinfo {title} {Density wave and supersolid phases of correlated
  bosons in an optical lattice},\ }\href@noop {} {\bibfield  {journal}
  {\bibinfo  {journal} {Europhysics Letters}\ }\textbf {\bibinfo {volume}
  {72}},\ \bibinfo {pages} {162} (\bibinfo {year} {2005})}\BibitemShut
  {NoStop}%
\bibitem [{\citenamefont {Shimizu}\ \emph {et~al.}(2018)\citenamefont
  {Shimizu}, \citenamefont {Hirano}, \citenamefont {Park}, \citenamefont
  {Kuno},\ and\ \citenamefont {Ichinose}}]{shimizu2018dynamics}%
  \BibitemOpen
  \bibfield  {author} {\bibinfo {author} {\bibfnamefont {K.}~\bibnamefont
  {Shimizu}}, \bibinfo {author} {\bibfnamefont {T.}~\bibnamefont {Hirano}},
  \bibinfo {author} {\bibfnamefont {J.}~\bibnamefont {Park}}, \bibinfo {author}
  {\bibfnamefont {Y.}~\bibnamefont {Kuno}},\ and\ \bibinfo {author}
  {\bibfnamefont {I.}~\bibnamefont {Ichinose}},\ }\bibfield  {title} {\bibinfo
  {title} {Dynamics of first-order quantum phase transitions in extended
  bose--hubbard model: from density wave to superfluid and vice versa},\
  }\href@noop {} {\bibfield  {journal} {\bibinfo  {journal} {New Journal of
  Physics}\ }\textbf {\bibinfo {volume} {20}},\ \bibinfo {pages} {083006}
  (\bibinfo {year} {2018})}\BibitemShut {NoStop}%
\bibitem [{\citenamefont {Sachdev}(2001)}]{QPTBook2001}%
  \BibitemOpen
  \bibfield  {author} {\bibinfo {author} {\bibfnamefont {S.}~\bibnamefont
  {Sachdev}},\ }\href@noop {} {\emph {\bibinfo {title} {Quantum {P}hase
  {T}ransitions}}}\ (\bibinfo  {publisher} {Cambridge University Press},\
  \bibinfo {year} {2001})\BibitemShut {NoStop}%
\bibitem [{\citenamefont {Tasgin}(2017)}]{tasgin2017many}%
  \BibitemOpen
  \bibfield  {author} {\bibinfo {author} {\bibfnamefont {M.~E.}\ \bibnamefont
  {Tasgin}},\ }\bibfield  {title} {\bibinfo {title} {Many-particle entanglement
  criterion for superradiantlike states},\ }\href@noop {} {\bibfield  {journal}
  {\bibinfo  {journal} {Physical Review Letters}\ }\textbf {\bibinfo {volume}
  {119}},\ \bibinfo {pages} {033601} (\bibinfo {year} {2017})}\BibitemShut
  {NoStop}%
\bibitem [{\citenamefont {Lambert}\ \emph {et~al.}(2004)\citenamefont
  {Lambert}, \citenamefont {Emary},\ and\ \citenamefont
  {Brandes}}]{lambert2004entanglement}%
  \BibitemOpen
  \bibfield  {author} {\bibinfo {author} {\bibfnamefont {N.}~\bibnamefont
  {Lambert}}, \bibinfo {author} {\bibfnamefont {C.}~\bibnamefont {Emary}},\
  and\ \bibinfo {author} {\bibfnamefont {T.}~\bibnamefont {Brandes}},\
  }\bibfield  {title} {\bibinfo {title} {Entanglement and the phase transition
  in single-mode superradiance},\ }\href@noop {} {\bibfield  {journal}
  {\bibinfo  {journal} {Physical Review Letters}\ }\textbf {\bibinfo {volume}
  {92}},\ \bibinfo {pages} {073602} (\bibinfo {year} {2004})}\BibitemShut
  {NoStop}%
\bibitem [{\citenamefont {G{\"u}nay}\ \emph {et~al.}(2019)\citenamefont
  {G{\"u}nay}, \citenamefont {M{\"u}stecapl{\i}o{\u{g}}lu},\ and\ \citenamefont
  {Tasgin}}]{gunay2019entanglement}%
  \BibitemOpen
  \bibfield  {author} {\bibinfo {author} {\bibfnamefont {M.}~\bibnamefont
  {G{\"u}nay}}, \bibinfo {author} {\bibfnamefont {{\"O}.~E.}\ \bibnamefont
  {M{\"u}stecapl{\i}o{\u{g}}lu}},\ and\ \bibinfo {author} {\bibfnamefont
  {M.~E.}\ \bibnamefont {Tasgin}},\ }\bibfield  {title} {\bibinfo {title}
  {Entanglement criteria for two strongly interacting ensembles},\ }\href@noop
  {} {\bibfield  {journal} {\bibinfo  {journal} {Physical Review A}\ }\textbf
  {\bibinfo {volume} {100}},\ \bibinfo {pages} {063838} (\bibinfo {year}
  {2019})}\BibitemShut {NoStop}%
\bibitem [{\citenamefont {G{\"u}nay}\ \emph {et~al.}(2023)\citenamefont
  {G{\"u}nay}, \citenamefont {Das}, \citenamefont {Y{\"u}ce}, \citenamefont
  {Polat}, \citenamefont {Bek},\ and\ \citenamefont
  {Tasgin}}]{gunay2023demand}%
  \BibitemOpen
  \bibfield  {author} {\bibinfo {author} {\bibfnamefont {M.}~\bibnamefont
  {G{\"u}nay}}, \bibinfo {author} {\bibfnamefont {P.}~\bibnamefont {Das}},
  \bibinfo {author} {\bibfnamefont {E.}~\bibnamefont {Y{\"u}ce}}, \bibinfo
  {author} {\bibfnamefont {E.~O.}\ \bibnamefont {Polat}}, \bibinfo {author}
  {\bibfnamefont {A.}~\bibnamefont {Bek}},\ and\ \bibinfo {author}
  {\bibfnamefont {M.~E.}\ \bibnamefont {Tasgin}},\ }\bibfield  {title}
  {\bibinfo {title} {On-demand continuous-variable quantum entanglement source
  for integrated circuits},\ }\href@noop {} {\bibfield  {journal} {\bibinfo
  {journal} {Nanophotonics}\ }\textbf {\bibinfo {volume} {12}},\ \bibinfo
  {pages} {229} (\bibinfo {year} {2023})}\BibitemShut {NoStop}%
\bibitem [{\citenamefont {Enomoto}\ \emph {et~al.}(2021)\citenamefont
  {Enomoto}, \citenamefont {Yonezu}, \citenamefont {Mitsuhashi}, \citenamefont
  {Takase},\ and\ \citenamefont {Takeda}}]{enomoto2021programmable}%
  \BibitemOpen
  \bibfield  {author} {\bibinfo {author} {\bibfnamefont {Y.}~\bibnamefont
  {Enomoto}}, \bibinfo {author} {\bibfnamefont {K.}~\bibnamefont {Yonezu}},
  \bibinfo {author} {\bibfnamefont {Y.}~\bibnamefont {Mitsuhashi}}, \bibinfo
  {author} {\bibfnamefont {K.}~\bibnamefont {Takase}},\ and\ \bibinfo {author}
  {\bibfnamefont {S.}~\bibnamefont {Takeda}},\ }\bibfield  {title} {\bibinfo
  {title} {Programmable and sequential gaussian gates in a loop-based
  single-mode photonic quantum processor},\ }\href@noop {} {\bibfield
  {journal} {\bibinfo  {journal} {Science Advances}\ }\textbf {\bibinfo
  {volume} {7}},\ \bibinfo {pages} {eabj6624} (\bibinfo {year}
  {2021})}\BibitemShut {NoStop}%
\bibitem [{\citenamefont {Ullah}\ \emph {et~al.}(2023)\citenamefont {Ullah},
  \citenamefont {Tasgin}, \citenamefont {Ovali},\ and\ \citenamefont
  {G{\"u}nay}}]{ullah2023electrically}%
  \BibitemOpen
  \bibfield  {author} {\bibinfo {author} {\bibfnamefont {S.}~\bibnamefont
  {Ullah}}, \bibinfo {author} {\bibfnamefont {M.~E.}\ \bibnamefont {Tasgin}},
  \bibinfo {author} {\bibfnamefont {R.~V.}\ \bibnamefont {Ovali}},\ and\
  \bibinfo {author} {\bibfnamefont {M.}~\bibnamefont {G{\"u}nay}},\ }\bibfield
  {title} {\bibinfo {title} {Electrically-programmable frequency comb for
  compact quantum photonic circuits},\ }\href@noop {} {\bibfield  {journal}
  {\bibinfo  {journal} {arXiv preprint arXiv:2308.00439}\ } (\bibinfo {year}
  {2023})}\BibitemShut {NoStop}%
\bibitem [{opt(2023)}]{optical_lattice_QC}%
  \BibitemOpen
  \href@noop {} {\bibinfo {title} {{M}ilestone for {O}ptical-{L}attice
  {Q}uantum {C}somputer}} (\bibinfo {year} {2023}),\ \bibinfo {note}
  {{https://physics.aps.org/articles/v16/s122}}\BibitemShut {NoStop}%
\bibitem [{Note2()}]{Note2}%
  \BibitemOpen
  \bibinfo {note} {For instance, in a superradiant or superfluid phase, the
  quasiparticle excitations exhibit number-squeezed correlation while in a
  spin-squeezed ensemble they show quadrature squeezed correlation~\cite
  {tasgin2017many}. Photon number states are known to possess nonclassicality
  stronger than the quadrature squeezed states.}\BibitemShut {Stop}%
\bibitem [{\citenamefont {Fr{\'e}rot}\ and\ \citenamefont
  {Roscilde}(2016)}]{frerot2016entanglement}%
  \BibitemOpen
  \bibfield  {author} {\bibinfo {author} {\bibfnamefont {I.}~\bibnamefont
  {Fr{\'e}rot}}\ and\ \bibinfo {author} {\bibfnamefont {T.}~\bibnamefont
  {Roscilde}},\ }\bibfield  {title} {\bibinfo {title} {Entanglement entropy
  across the superfluid-insulator transition: A signature of bosonic
  criticality},\ }\href@noop {} {\bibfield  {journal} {\bibinfo  {journal}
  {Physical Review Letters}\ }\textbf {\bibinfo {volume} {116}},\ \bibinfo
  {pages} {190401} (\bibinfo {year} {2016})}\BibitemShut {NoStop}%
\bibitem [{\citenamefont {Walsh}\ \emph {et~al.}(2019)\citenamefont {Walsh},
  \citenamefont {S{\'e}mon}, \citenamefont {Poulin}, \citenamefont {Sordi},\
  and\ \citenamefont {Tremblay}}]{walsh2019local}%
  \BibitemOpen
  \bibfield  {author} {\bibinfo {author} {\bibfnamefont {C.}~\bibnamefont
  {Walsh}}, \bibinfo {author} {\bibfnamefont {P.}~\bibnamefont {S{\'e}mon}},
  \bibinfo {author} {\bibfnamefont {D.}~\bibnamefont {Poulin}}, \bibinfo
  {author} {\bibfnamefont {G.}~\bibnamefont {Sordi}},\ and\ \bibinfo {author}
  {\bibfnamefont {A.-M.}\ \bibnamefont {Tremblay}},\ }\bibfield  {title}
  {\bibinfo {title} {Local entanglement entropy and mutual information across
  the mott transition in the two-dimensional hubbard model},\ }\href@noop {}
  {\bibfield  {journal} {\bibinfo  {journal} {Physical Review Letters}\
  }\textbf {\bibinfo {volume} {122}},\ \bibinfo {pages} {067203} (\bibinfo
  {year} {2019})}\BibitemShut {NoStop}%
\bibitem [{\citenamefont {Sharma}\ \emph {et~al.}(2022)\citenamefont {Sharma},
  \citenamefont {J{\"a}ger}, \citenamefont {Kraus}, \citenamefont {Roscilde},\
  and\ \citenamefont {Morigi}}]{sharma2022quantum}%
  \BibitemOpen
  \bibfield  {author} {\bibinfo {author} {\bibfnamefont {S.}~\bibnamefont
  {Sharma}}, \bibinfo {author} {\bibfnamefont {S.~B.}\ \bibnamefont
  {J{\"a}ger}}, \bibinfo {author} {\bibfnamefont {R.}~\bibnamefont {Kraus}},
  \bibinfo {author} {\bibfnamefont {T.}~\bibnamefont {Roscilde}},\ and\
  \bibinfo {author} {\bibfnamefont {G.}~\bibnamefont {Morigi}},\ }\bibfield
  {title} {\bibinfo {title} {Quantum critical behavior of entanglement in
  lattice bosons with cavity-mediated long-range interactions},\ }\href@noop {}
  {\bibfield  {journal} {\bibinfo  {journal} {Physical Review Letters}\
  }\textbf {\bibinfo {volume} {129}},\ \bibinfo {pages} {143001} (\bibinfo
  {year} {2022})}\BibitemShut {NoStop}%
\bibitem [{\citenamefont {Zhang}\ \emph {et~al.}(2016)\citenamefont {Zhang},
  \citenamefont {Qin}, \citenamefont {Ke}, \citenamefont {Lee} \emph
  {et~al.}}]{zhang2016cluster}%
  \BibitemOpen
  \bibfield  {author} {\bibinfo {author} {\bibfnamefont {L.}~\bibnamefont
  {Zhang}}, \bibinfo {author} {\bibfnamefont {X.}~\bibnamefont {Qin}}, \bibinfo
  {author} {\bibfnamefont {Y.}~\bibnamefont {Ke}}, \bibinfo {author}
  {\bibfnamefont {C.}~\bibnamefont {Lee}}, \emph {et~al.},\ }\bibfield  {title}
  {\bibinfo {title} {Cluster mean-field signature of entanglement entropy in
  bosonic superfluid-insulator transitions},\ }\href@noop {} {\bibfield
  {journal} {\bibinfo  {journal} {Physical Review A}\ }\textbf {\bibinfo
  {volume} {94}},\ \bibinfo {pages} {023634} (\bibinfo {year}
  {2016})}\BibitemShut {NoStop}%
\bibitem [{\citenamefont {Canella}\ and\ \citenamefont
  {Fran{\c{c}}a}(2019)}]{canella2019superfluid}%
  \BibitemOpen
  \bibfield  {author} {\bibinfo {author} {\bibfnamefont {G.~A.}\ \bibnamefont
  {Canella}}\ and\ \bibinfo {author} {\bibfnamefont {V.~V.}\ \bibnamefont
  {Fran{\c{c}}a}},\ }\bibfield  {title} {\bibinfo {title} {Superfluid-insulator
  transition unambiguously detected by entanglement in one-dimensional
  disordered superfluids},\ }\href@noop {} {\bibfield  {journal} {\bibinfo
  {journal} {Scientific Reports}\ }\textbf {\bibinfo {volume} {9}},\ \bibinfo
  {pages} {15313} (\bibinfo {year} {2019})}\BibitemShut {NoStop}%
\bibitem [{\citenamefont {Maschler}\ \emph {et~al.}(2008)\citenamefont
  {Maschler}, \citenamefont {Mekhov},\ and\ \citenamefont
  {Ritsch}}]{ritsch2008ultracold}%
  \BibitemOpen
  \bibfield  {author} {\bibinfo {author} {\bibfnamefont {C.}~\bibnamefont
  {Maschler}}, \bibinfo {author} {\bibfnamefont {I.~B.}\ \bibnamefont
  {Mekhov}},\ and\ \bibinfo {author} {\bibfnamefont {H.}~\bibnamefont
  {Ritsch}},\ }\bibfield  {title} {\bibinfo {title} {Ultracold atoms in optical
  lattices generated by quantized light fields},\ }\href@noop {} {\bibfield
  {journal} {\bibinfo  {journal} {The European Physical Journal D}\ }\textbf
  {\bibinfo {volume} {46}},\ \bibinfo {pages} {545} (\bibinfo {year}
  {2008})}\BibitemShut {NoStop}%
\bibitem [{Note3()}]{Note3}%
  \BibitemOpen
  \bibinfo {note} {D. F. V. James {\protect \it et al.}, Phys.\ Rev.\
  A{\protect \bf 64}, 052312 (2001).}\BibitemShut {Stop}%
\bibitem [{Note4()}]{Note4}%
  \BibitemOpen
  \bibinfo {note} {R. Islam {\protect \it et al.}, Nature {\protect \bf 528},
  77 (2015).}\BibitemShut {Stop}%
\bibitem [{Note5()}]{Note5}%
  \BibitemOpen
  \bibinfo {note} {T. Brydges {\protect \it et al.} Science {\protect \bf 364},
  260 (2019).}\BibitemShut {Stop}%
\bibitem [{PS_()}]{PS_crossover}%
  \BibitemOpen
  \href@noop {} {}\bibinfo {note} {Here, we prefer to use the word crossover as
  our exact simulations are with very few number of particles/sites, i.e.,
  $L=N=8$.}\BibitemShut {Stop}%
\bibitem [{\citenamefont {Dao}\ \emph {et~al.}(2007)\citenamefont {Dao},
  \citenamefont {Georges}, \citenamefont {Dalibard}, \citenamefont {Salomon},\
  and\ \citenamefont {Carusotto}}]{dao2007measuring}%
  \BibitemOpen
  \bibfield  {author} {\bibinfo {author} {\bibfnamefont {T.-L.}\ \bibnamefont
  {Dao}}, \bibinfo {author} {\bibfnamefont {A.}~\bibnamefont {Georges}},
  \bibinfo {author} {\bibfnamefont {J.}~\bibnamefont {Dalibard}}, \bibinfo
  {author} {\bibfnamefont {C.}~\bibnamefont {Salomon}},\ and\ \bibinfo {author}
  {\bibfnamefont {I.}~\bibnamefont {Carusotto}},\ }\bibfield  {title} {\bibinfo
  {title} {Measuring the one-particle excitations of ultracold fermionic atoms
  by stimulated raman spectroscopy},\ }\href@noop {} {\bibfield  {journal}
  {\bibinfo  {journal} {Physical Review Letters}\ }\textbf {\bibinfo {volume}
  {98}},\ \bibinfo {pages} {240402} (\bibinfo {year} {2007})}\BibitemShut
  {NoStop}%
\bibitem [{\citenamefont {Blakie}(2006)}]{blakie2006raman}%
  \BibitemOpen
  \bibfield  {author} {\bibinfo {author} {\bibfnamefont {P.~B.}\ \bibnamefont
  {Blakie}},\ }\bibfield  {title} {\bibinfo {title} {Raman spectroscopy of mott
  insulator states in optical lattices},\ }\href@noop {} {\bibfield  {journal}
  {\bibinfo  {journal} {New Journal of Physics}\ }\textbf {\bibinfo {volume}
  {8}},\ \bibinfo {pages} {157} (\bibinfo {year} {2006})}\BibitemShut {NoStop}%
\bibitem [{\citenamefont {Konabe}\ \emph {et~al.}(2006)\citenamefont {Konabe},
  \citenamefont {Nikuni},\ and\ \citenamefont {Nakamura}}]{konabe2006laser}%
  \BibitemOpen
  \bibfield  {author} {\bibinfo {author} {\bibfnamefont {S.}~\bibnamefont
  {Konabe}}, \bibinfo {author} {\bibfnamefont {T.}~\bibnamefont {Nikuni}},\
  and\ \bibinfo {author} {\bibfnamefont {M.}~\bibnamefont {Nakamura}},\
  }\bibfield  {title} {\bibinfo {title} {Laser probing of the single-particle
  energy gap of a bose gas in an optical lattice in the mott-insulator phase},\
  }\href@noop {} {\bibfield  {journal} {\bibinfo  {journal} {Physical Review
  A}\ }\textbf {\bibinfo {volume} {73}},\ \bibinfo {pages} {033621} (\bibinfo
  {year} {2006})}\BibitemShut {NoStop}%
\bibitem [{\citenamefont {Dao}\ \emph {et~al.}(2009)\citenamefont {Dao},
  \citenamefont {Carusotto},\ and\ \citenamefont {Georges}}]{dao2009probing}%
  \BibitemOpen
  \bibfield  {author} {\bibinfo {author} {\bibfnamefont {T.-L.}\ \bibnamefont
  {Dao}}, \bibinfo {author} {\bibfnamefont {I.}~\bibnamefont {Carusotto}},\
  and\ \bibinfo {author} {\bibfnamefont {A.}~\bibnamefont {Georges}},\
  }\bibfield  {title} {\bibinfo {title} {Probing quasiparticle states in
  strongly interacting atomic gases by momentum-resolved raman photoemission
  spectroscopy},\ }\href@noop {} {\bibfield  {journal} {\bibinfo  {journal}
  {Physical Review A}\ }\textbf {\bibinfo {volume} {80}},\ \bibinfo {pages}
  {023627} (\bibinfo {year} {2009})}\BibitemShut {NoStop}%
\bibitem [{\citenamefont {Leskinen}\ \emph {et~al.}(2010)\citenamefont
  {Leskinen}, \citenamefont {Kajala},\ and\ \citenamefont
  {Kinnunen}}]{leskinen2010resonant}%
  \BibitemOpen
  \bibfield  {author} {\bibinfo {author} {\bibfnamefont {M.}~\bibnamefont
  {Leskinen}}, \bibinfo {author} {\bibfnamefont {J.}~\bibnamefont {Kajala}},\
  and\ \bibinfo {author} {\bibfnamefont {J.}~\bibnamefont {Kinnunen}},\
  }\bibfield  {title} {\bibinfo {title} {Resonant scattering effect in
  spectroscopies of interacting atomic gases},\ }\href@noop {} {\bibfield
  {journal} {\bibinfo  {journal} {New Journal of Physics}\ }\textbf {\bibinfo
  {volume} {12}},\ \bibinfo {pages} {083041} (\bibinfo {year}
  {2010})}\BibitemShut {NoStop}%
\bibitem [{\citenamefont {Hall}\ \emph {et~al.}(1998)\citenamefont {Hall},
  \citenamefont {Matthews}, \citenamefont {Ensher}, \citenamefont {Wieman},\
  and\ \citenamefont {Cornell}}]{hall1998dynamics}%
  \BibitemOpen
  \bibfield  {author} {\bibinfo {author} {\bibfnamefont {D.}~\bibnamefont
  {Hall}}, \bibinfo {author} {\bibfnamefont {M.}~\bibnamefont {Matthews}},
  \bibinfo {author} {\bibfnamefont {J.}~\bibnamefont {Ensher}}, \bibinfo
  {author} {\bibfnamefont {C.}~\bibnamefont {Wieman}},\ and\ \bibinfo {author}
  {\bibfnamefont {E.~A.}\ \bibnamefont {Cornell}},\ }\bibfield  {title}
  {\bibinfo {title} {Dynamics of component separation in a binary mixture of
  bose-einstein condensates},\ }\href@noop {} {\bibfield  {journal} {\bibinfo
  {journal} {Physical Review Letters}\ }\textbf {\bibinfo {volume} {81}},\
  \bibinfo {pages} {1539} (\bibinfo {year} {1998})}\BibitemShut {NoStop}%
\bibitem [{\citenamefont {Bernier}\ \emph {et~al.}(2010)\citenamefont
  {Bernier}, \citenamefont {Dao}, \citenamefont {Kollath}, \citenamefont
  {Georges},\ and\ \citenamefont {Cornaglia}}]{bernier2010thermometry}%
  \BibitemOpen
  \bibfield  {author} {\bibinfo {author} {\bibfnamefont {J.-S.}\ \bibnamefont
  {Bernier}}, \bibinfo {author} {\bibfnamefont {T.-L.}\ \bibnamefont {Dao}},
  \bibinfo {author} {\bibfnamefont {C.}~\bibnamefont {Kollath}}, \bibinfo
  {author} {\bibfnamefont {A.}~\bibnamefont {Georges}},\ and\ \bibinfo {author}
  {\bibfnamefont {P.~S.}\ \bibnamefont {Cornaglia}},\ }\bibfield  {title}
  {\bibinfo {title} {Thermometry and signatures of strong correlations from
  raman spectroscopy of fermionic atoms in optical lattices},\ }\href@noop {}
  {\bibfield  {journal} {\bibinfo  {journal} {Physical Review A}\ }\textbf
  {\bibinfo {volume} {81}},\ \bibinfo {pages} {063618} (\bibinfo {year}
  {2010})}\BibitemShut {NoStop}%
\bibitem [{\citenamefont {Kozlowski}\ \emph {et~al.}(2017)\citenamefont
  {Kozlowski}, \citenamefont {Caballero-Benitez},\ and\ \citenamefont
  {Mekhov}}]{kozlowski2017quantum}%
  \BibitemOpen
  \bibfield  {author} {\bibinfo {author} {\bibfnamefont {W.}~\bibnamefont
  {Kozlowski}}, \bibinfo {author} {\bibfnamefont {S.~F.}\ \bibnamefont
  {Caballero-Benitez}},\ and\ \bibinfo {author} {\bibfnamefont {I.~B.}\
  \bibnamefont {Mekhov}},\ }\bibfield  {title} {\bibinfo {title} {Quantum state
  reduction by matter-phase-related measurements in optical lattices},\
  }\href@noop {} {\bibfield  {journal} {\bibinfo  {journal} {Scientific
  Reports}\ }\textbf {\bibinfo {volume} {7}},\ \bibinfo {pages} {42597}
  (\bibinfo {year} {2017})}\BibitemShut {NoStop}%
\bibitem [{\citenamefont {Mekhov}\ \emph {et~al.}(2007)\citenamefont {Mekhov},
  \citenamefont {Maschler},\ and\ \citenamefont {Ritsch}}]{mekhov2007light}%
  \BibitemOpen
  \bibfield  {author} {\bibinfo {author} {\bibfnamefont {I.~B.}\ \bibnamefont
  {Mekhov}}, \bibinfo {author} {\bibfnamefont {C.}~\bibnamefont {Maschler}},\
  and\ \bibinfo {author} {\bibfnamefont {H.}~\bibnamefont {Ritsch}},\
  }\bibfield  {title} {\bibinfo {title} {Light scattering from ultracold atoms
  in optical lattices as an optical probe of quantum statistics},\ }\href@noop
  {} {\bibfield  {journal} {\bibinfo  {journal} {Physical Review A}\ }\textbf
  {\bibinfo {volume} {76}},\ \bibinfo {pages} {053618} (\bibinfo {year}
  {2007})}\BibitemShut {NoStop}%
\bibitem [{\citenamefont {Kozlowski}\ \emph {et~al.}(2015)\citenamefont
  {Kozlowski}, \citenamefont {Caballero-Benitez},\ and\ \citenamefont
  {Mekhov}}]{kozlowski2015probing}%
  \BibitemOpen
  \bibfield  {author} {\bibinfo {author} {\bibfnamefont {W.}~\bibnamefont
  {Kozlowski}}, \bibinfo {author} {\bibfnamefont {S.~F.}\ \bibnamefont
  {Caballero-Benitez}},\ and\ \bibinfo {author} {\bibfnamefont {I.~B.}\
  \bibnamefont {Mekhov}},\ }\bibfield  {title} {\bibinfo {title} {Probing
  matter-field and atom-number correlations in optical lattices by global
  nondestructive addressing},\ }\href@noop {} {\bibfield  {journal} {\bibinfo
  {journal} {Physical Review A}\ }\textbf {\bibinfo {volume} {92}},\ \bibinfo
  {pages} {013613} (\bibinfo {year} {2015})}\BibitemShut {NoStop}%
\bibitem [{\citenamefont {Javanainen}\ and\ \citenamefont
  {Ruostekoski}(1995)}]{javanainen1995off}%
  \BibitemOpen
  \bibfield  {author} {\bibinfo {author} {\bibfnamefont {J.}~\bibnamefont
  {Javanainen}}\ and\ \bibinfo {author} {\bibfnamefont {J.}~\bibnamefont
  {Ruostekoski}},\ }\bibfield  {title} {\bibinfo {title} {Off-resonance light
  scattering from low-temperature bose and fermi gases},\ }\href@noop {}
  {\bibfield  {journal} {\bibinfo  {journal} {Physical Review A}\ }\textbf
  {\bibinfo {volume} {52}},\ \bibinfo {pages} {3033} (\bibinfo {year}
  {1995})}\BibitemShut {NoStop}%
\bibitem [{\citenamefont {Gerbier}\ \emph {et~al.}(2008)\citenamefont
  {Gerbier}, \citenamefont {Trotzky}, \citenamefont {F{\"o}lling},
  \citenamefont {Schnorrberger}, \citenamefont {Thompson}, \citenamefont
  {Widera}, \citenamefont {Bloch}, \citenamefont {Pollet}, \citenamefont
  {Troyer}, \citenamefont {Capogrosso-Sansone} \emph
  {et~al.}}]{gerbier2008expansion}%
  \BibitemOpen
  \bibfield  {author} {\bibinfo {author} {\bibfnamefont {F.}~\bibnamefont
  {Gerbier}}, \bibinfo {author} {\bibfnamefont {S.}~\bibnamefont {Trotzky}},
  \bibinfo {author} {\bibfnamefont {S.}~\bibnamefont {F{\"o}lling}}, \bibinfo
  {author} {\bibfnamefont {U.}~\bibnamefont {Schnorrberger}}, \bibinfo {author}
  {\bibfnamefont {J.}~\bibnamefont {Thompson}}, \bibinfo {author}
  {\bibfnamefont {A.}~\bibnamefont {Widera}}, \bibinfo {author} {\bibfnamefont
  {I.}~\bibnamefont {Bloch}}, \bibinfo {author} {\bibfnamefont
  {L.}~\bibnamefont {Pollet}}, \bibinfo {author} {\bibfnamefont
  {M.}~\bibnamefont {Troyer}}, \bibinfo {author} {\bibfnamefont
  {B.}~\bibnamefont {Capogrosso-Sansone}}, \emph {et~al.},\ }\bibfield  {title}
  {\bibinfo {title} {Expansion of a quantum gas released from an optical
  lattice},\ }\href@noop {} {\bibfield  {journal} {\bibinfo  {journal}
  {Physical Review Letters}\ }\textbf {\bibinfo {volume} {101}},\ \bibinfo
  {pages} {155303} (\bibinfo {year} {2008})}\BibitemShut {NoStop}%
\bibitem [{\citenamefont {S{\o}rensen}\ \emph {et~al.}(2001)\citenamefont
  {S{\o}rensen}, \citenamefont {Duan}, \citenamefont {Cirac},\ and\
  \citenamefont {Zoller}}]{sorensen2001Nature}%
  \BibitemOpen
  \bibfield  {author} {\bibinfo {author} {\bibfnamefont {A.}~\bibnamefont
  {S{\o}rensen}}, \bibinfo {author} {\bibfnamefont {L.-M.}\ \bibnamefont
  {Duan}}, \bibinfo {author} {\bibfnamefont {J.}~\bibnamefont {Cirac}},\ and\
  \bibinfo {author} {\bibfnamefont {P.}~\bibnamefont {Zoller}},\ }\bibfield
  {title} {\bibinfo {title} {Many-particle entanglement with bose--einstein
  condensates},\ }\href@noop {} {\bibfield  {journal} {\bibinfo  {journal}
  {Nature}\ }\textbf {\bibinfo {volume} {409}},\ \bibinfo {pages} {63}
  (\bibinfo {year} {2001})}\BibitemShut {NoStop}%
\bibitem [{\citenamefont {Duan}\ \emph {et~al.}(2000)\citenamefont {Duan},
  \citenamefont {Giedke}, \citenamefont {Cirac},\ and\ \citenamefont
  {Zoller}}]{DGCZ_PRL2000}%
  \BibitemOpen
  \bibfield  {author} {\bibinfo {author} {\bibfnamefont {L.-M.}\ \bibnamefont
  {Duan}}, \bibinfo {author} {\bibfnamefont {G.}~\bibnamefont {Giedke}},
  \bibinfo {author} {\bibfnamefont {J.}~\bibnamefont {Cirac}},\ and\ \bibinfo
  {author} {\bibfnamefont {P.}~\bibnamefont {Zoller}},\ }\bibfield  {title}
  {\bibinfo {title} {Inseparability criterion for continuous variable
  systems},\ }\href {https://doi.org/10.1103/PhysRevLett.84.2722} {\bibfield
  {journal} {\bibinfo  {journal} {Phys. Rev. Lett.}\ }\textbf {\bibinfo
  {volume} {84}},\ \bibinfo {pages} {2722} (\bibinfo {year}
  {2000})}\BibitemShut {NoStop}%
\bibitem [{\citenamefont {Roy}\ \emph {et~al.}(2022)\citenamefont {Roy},
  \citenamefont {Carl},\ and\ \citenamefont {Hauke}}]{roy2022genuine}%
  \BibitemOpen
  \bibfield  {author} {\bibinfo {author} {\bibfnamefont {S.~S.}\ \bibnamefont
  {Roy}}, \bibinfo {author} {\bibfnamefont {L.}~\bibnamefont {Carl}},\ and\
  \bibinfo {author} {\bibfnamefont {P.}~\bibnamefont {Hauke}},\ }\bibfield
  {title} {\bibinfo {title} {Genuine multipartite entanglement in a
  one-dimensional bose-hubbard model with frustrated hopping},\ }\href@noop {}
  {\bibfield  {journal} {\bibinfo  {journal} {Physical Review B}\ }\textbf
  {\bibinfo {volume} {106}},\ \bibinfo {pages} {195158} (\bibinfo {year}
  {2022})}\BibitemShut {NoStop}%
\bibitem [{sup()}]{supplementary}%
  \BibitemOpen
  \href@noop {} {}\bibinfo {note} {See Supplemental Material at
  http://link.aps.org/ supplemental/}\BibitemShut {NoStop}%
\bibitem [{\citenamefont {Duan}(2011)}]{duan2011many_particle_entanglement}%
  \BibitemOpen
  \bibfield  {author} {\bibinfo {author} {\bibfnamefont {L.-M.}\ \bibnamefont
  {Duan}},\ }\bibfield  {title} {\bibinfo {title} {Entanglement detection in
  the vicinity of arbitrary dicke states},\ }\href@noop {} {\bibfield
  {journal} {\bibinfo  {journal} {Physical Review Letters}\ }\textbf {\bibinfo
  {volume} {107}},\ \bibinfo {pages} {180502} (\bibinfo {year}
  {2011})}\BibitemShut {NoStop}%
\bibitem [{\citenamefont {Bakr}\ \emph {et~al.}(2009)\citenamefont {Bakr},
  \citenamefont {Gillen}, \citenamefont {Peng}, \citenamefont {F{\"o}lling},\
  and\ \citenamefont {Greiner}}]{bakr2009quantum}%
  \BibitemOpen
  \bibfield  {author} {\bibinfo {author} {\bibfnamefont {W.~S.}\ \bibnamefont
  {Bakr}}, \bibinfo {author} {\bibfnamefont {J.~I.}\ \bibnamefont {Gillen}},
  \bibinfo {author} {\bibfnamefont {A.}~\bibnamefont {Peng}}, \bibinfo {author}
  {\bibfnamefont {S.}~\bibnamefont {F{\"o}lling}},\ and\ \bibinfo {author}
  {\bibfnamefont {M.}~\bibnamefont {Greiner}},\ }\bibfield  {title} {\bibinfo
  {title} {A quantum gas microscope for detecting single atoms in a
  hubbard-regime optical lattice},\ }\href@noop {} {\bibfield  {journal}
  {\bibinfo  {journal} {Nature}\ }\textbf {\bibinfo {volume} {462}},\ \bibinfo
  {pages} {74} (\bibinfo {year} {2009})}\BibitemShut {NoStop}%
\bibitem [{\citenamefont {Sherson}\ \emph {et~al.}(2010)\citenamefont
  {Sherson}, \citenamefont {Weitenberg}, \citenamefont {Endres}, \citenamefont
  {Cheneau}, \citenamefont {Bloch},\ and\ \citenamefont
  {Kuhr}}]{sherson2010single}%
  \BibitemOpen
  \bibfield  {author} {\bibinfo {author} {\bibfnamefont {J.~F.}\ \bibnamefont
  {Sherson}}, \bibinfo {author} {\bibfnamefont {C.}~\bibnamefont {Weitenberg}},
  \bibinfo {author} {\bibfnamefont {M.}~\bibnamefont {Endres}}, \bibinfo
  {author} {\bibfnamefont {M.}~\bibnamefont {Cheneau}}, \bibinfo {author}
  {\bibfnamefont {I.}~\bibnamefont {Bloch}},\ and\ \bibinfo {author}
  {\bibfnamefont {S.}~\bibnamefont {Kuhr}},\ }\bibfield  {title} {\bibinfo
  {title} {Single-atom-resolved fluorescence imaging of an atomic mott
  insulator},\ }\href@noop {} {\bibfield  {journal} {\bibinfo  {journal}
  {Nature}\ }\textbf {\bibinfo {volume} {467}},\ \bibinfo {pages} {68}
  (\bibinfo {year} {2010})}\BibitemShut {NoStop}%
\bibitem [{\citenamefont {Campbell}\ \emph {et~al.}(2006)\citenamefont
  {Campbell}, \citenamefont {Mun}, \citenamefont {Boyd}, \citenamefont
  {Medley}, \citenamefont {Leanhardt}, \citenamefont {Marcassa}, \citenamefont
  {Pritchard},\ and\ \citenamefont {Ketterle}}]{campbell2006imaging}%
  \BibitemOpen
  \bibfield  {author} {\bibinfo {author} {\bibfnamefont {G.~K.}\ \bibnamefont
  {Campbell}}, \bibinfo {author} {\bibfnamefont {J.}~\bibnamefont {Mun}},
  \bibinfo {author} {\bibfnamefont {M.}~\bibnamefont {Boyd}}, \bibinfo {author}
  {\bibfnamefont {P.}~\bibnamefont {Medley}}, \bibinfo {author} {\bibfnamefont
  {A.~E.}\ \bibnamefont {Leanhardt}}, \bibinfo {author} {\bibfnamefont {L.~G.}\
  \bibnamefont {Marcassa}}, \bibinfo {author} {\bibfnamefont {D.~E.}\
  \bibnamefont {Pritchard}},\ and\ \bibinfo {author} {\bibfnamefont
  {W.}~\bibnamefont {Ketterle}},\ }\bibfield  {title} {\bibinfo {title}
  {Imaging the mott insulator shells by using atomic clock shifts},\
  }\href@noop {} {\bibfield  {journal} {\bibinfo  {journal} {Science}\ }\textbf
  {\bibinfo {volume} {313}},\ \bibinfo {pages} {649} (\bibinfo {year}
  {2006})}\BibitemShut {NoStop}%
\bibitem [{\citenamefont {Gerbier}\ \emph {et~al.}(2006)\citenamefont
  {Gerbier}, \citenamefont {F{\"o}lling}, \citenamefont {Widera}, \citenamefont
  {Mandel},\ and\ \citenamefont {Bloch}}]{gerbier2006probing}%
  \BibitemOpen
  \bibfield  {author} {\bibinfo {author} {\bibfnamefont {F.}~\bibnamefont
  {Gerbier}}, \bibinfo {author} {\bibfnamefont {S.}~\bibnamefont
  {F{\"o}lling}}, \bibinfo {author} {\bibfnamefont {A.}~\bibnamefont {Widera}},
  \bibinfo {author} {\bibfnamefont {O.}~\bibnamefont {Mandel}},\ and\ \bibinfo
  {author} {\bibfnamefont {I.}~\bibnamefont {Bloch}},\ }\bibfield  {title}
  {\bibinfo {title} {Probing number squeezing of ultracold atoms across the
  superfluid-mott insulator transition},\ }\href@noop {} {\bibfield  {journal}
  {\bibinfo  {journal} {Physical review letters}\ }\textbf {\bibinfo {volume}
  {96}},\ \bibinfo {pages} {090401} (\bibinfo {year} {2006})}\BibitemShut
  {NoStop}%
\bibitem [{\citenamefont {Kele{\c{s}}}\ and\ \citenamefont
  {Oktel}(2015)}]{kelecs2015mott}%
  \BibitemOpen
  \bibfield  {author} {\bibinfo {author} {\bibfnamefont {A.}~\bibnamefont
  {Kele{\c{s}}}}\ and\ \bibinfo {author} {\bibfnamefont {M.}~\bibnamefont
  {Oktel}},\ }\bibfield  {title} {\bibinfo {title} {Mott transition in a
  two-leg bose-hubbard ladder under an artificial magnetic field},\ }\href@noop
  {} {\bibfield  {journal} {\bibinfo  {journal} {Physical Review A}\ }\textbf
  {\bibinfo {volume} {91}},\ \bibinfo {pages} {013629} (\bibinfo {year}
  {2015})}\BibitemShut {NoStop}%
\bibitem [{\citenamefont {Cooper}\ \emph {et~al.}(2001)\citenamefont {Cooper},
  \citenamefont {Wilkin},\ and\ \citenamefont {Gunn}}]{cooper2001quantum}%
  \BibitemOpen
  \bibfield  {author} {\bibinfo {author} {\bibfnamefont {N.~R.}\ \bibnamefont
  {Cooper}}, \bibinfo {author} {\bibfnamefont {N.~K.}\ \bibnamefont {Wilkin}},\
  and\ \bibinfo {author} {\bibfnamefont {J.}~\bibnamefont {Gunn}},\ }\bibfield
  {title} {\bibinfo {title} {Quantum phases of vortices in rotating
  bose-einstein condensates},\ }\href@noop {} {\bibfield  {journal} {\bibinfo
  {journal} {Physical Review Letters}\ }\textbf {\bibinfo {volume} {87}},\
  \bibinfo {pages} {120405} (\bibinfo {year} {2001})}\BibitemShut {NoStop}%
\bibitem [{\citenamefont {Kozuma}\ \emph {et~al.}(1999)\citenamefont {Kozuma},
  \citenamefont {Deng}, \citenamefont {Hagley}, \citenamefont {Wen},
  \citenamefont {Lutwak}, \citenamefont {Helmerson}, \citenamefont {Rolston},\
  and\ \citenamefont {Phillips}}]{kozuma1999coherent}%
  \BibitemOpen
  \bibfield  {author} {\bibinfo {author} {\bibfnamefont {M.}~\bibnamefont
  {Kozuma}}, \bibinfo {author} {\bibfnamefont {L.}~\bibnamefont {Deng}},
  \bibinfo {author} {\bibfnamefont {E.~W.}\ \bibnamefont {Hagley}}, \bibinfo
  {author} {\bibfnamefont {J.}~\bibnamefont {Wen}}, \bibinfo {author}
  {\bibfnamefont {R.}~\bibnamefont {Lutwak}}, \bibinfo {author} {\bibfnamefont
  {K.}~\bibnamefont {Helmerson}}, \bibinfo {author} {\bibfnamefont
  {S.}~\bibnamefont {Rolston}},\ and\ \bibinfo {author} {\bibfnamefont {W.~D.}\
  \bibnamefont {Phillips}},\ }\bibfield  {title} {\bibinfo {title} {Coherent
  splitting of bose-einstein condensed atoms with optically induced bragg
  diffraction},\ }\href@noop {} {\bibfield  {journal} {\bibinfo  {journal}
  {Physical Review Letters}\ }\textbf {\bibinfo {volume} {82}},\ \bibinfo
  {pages} {871} (\bibinfo {year} {1999})}\BibitemShut {NoStop}%
\bibitem [{\citenamefont {Polkovnikov}\ \emph {et~al.}(2006)\citenamefont
  {Polkovnikov}, \citenamefont {Altman},\ and\ \citenamefont
  {Demler}}]{polkovnikov2006interference}%
  \BibitemOpen
  \bibfield  {author} {\bibinfo {author} {\bibfnamefont {A.}~\bibnamefont
  {Polkovnikov}}, \bibinfo {author} {\bibfnamefont {E.}~\bibnamefont
  {Altman}},\ and\ \bibinfo {author} {\bibfnamefont {E.}~\bibnamefont
  {Demler}},\ }\bibfield  {title} {\bibinfo {title} {Interference between
  independent fluctuating condensates},\ }\href@noop {} {\bibfield  {journal}
  {\bibinfo  {journal} {Proceedings of the National Academy of Sciences}\
  }\textbf {\bibinfo {volume} {103}},\ \bibinfo {pages} {6125} (\bibinfo {year}
  {2006})}\BibitemShut {NoStop}%
\bibitem [{\citenamefont {Niu}\ \emph {et~al.}(2006)\citenamefont {Niu},
  \citenamefont {Carusotto},\ and\ \citenamefont {Kuklov}}]{niu2006imaging}%
  \BibitemOpen
  \bibfield  {author} {\bibinfo {author} {\bibfnamefont {Q.}~\bibnamefont
  {Niu}}, \bibinfo {author} {\bibfnamefont {I.}~\bibnamefont {Carusotto}},\
  and\ \bibinfo {author} {\bibfnamefont {A.}~\bibnamefont {Kuklov}},\
  }\bibfield  {title} {\bibinfo {title} {Imaging of critical correlations in
  optical lattices and atomic traps},\ }\href@noop {} {\bibfield  {journal}
  {\bibinfo  {journal} {Physical Review A}\ }\textbf {\bibinfo {volume} {73}},\
  \bibinfo {pages} {053604} (\bibinfo {year} {2006})}\BibitemShut {NoStop}%
\bibitem [{\citenamefont {Jimenez-Garcia}\ \emph {et~al.}(2012)\citenamefont
  {Jimenez-Garcia}, \citenamefont {LeBlanc}, \citenamefont {Williams},
  \citenamefont {Beeler}, \citenamefont {Perry},\ and\ \citenamefont
  {Spielman}}]{jimenez2012peierls}%
  \BibitemOpen
  \bibfield  {author} {\bibinfo {author} {\bibfnamefont {K.}~\bibnamefont
  {Jimenez-Garcia}}, \bibinfo {author} {\bibfnamefont {L.~J.}\ \bibnamefont
  {LeBlanc}}, \bibinfo {author} {\bibfnamefont {R.~A.}\ \bibnamefont
  {Williams}}, \bibinfo {author} {\bibfnamefont {M.~C.}\ \bibnamefont
  {Beeler}}, \bibinfo {author} {\bibfnamefont {A.~R.}\ \bibnamefont {Perry}},\
  and\ \bibinfo {author} {\bibfnamefont {I.~B.}\ \bibnamefont {Spielman}},\
  }\bibfield  {title} {\bibinfo {title} {Peierls substitution in an engineered
  lattice potential},\ }\href@noop {} {\bibfield  {journal} {\bibinfo
  {journal} {Physical Review Letters}\ }\textbf {\bibinfo {volume} {108}},\
  \bibinfo {pages} {225303} (\bibinfo {year} {2012})}\BibitemShut {NoStop}%
\bibitem [{\citenamefont {Miyake}\ \emph {et~al.}(2011)\citenamefont {Miyake},
  \citenamefont {Siviloglou}, \citenamefont {Puentes}, \citenamefont
  {Pritchard}, \citenamefont {Ketterle},\ and\ \citenamefont
  {Weld}}]{miyake2011bragg}%
  \BibitemOpen
  \bibfield  {author} {\bibinfo {author} {\bibfnamefont {H.}~\bibnamefont
  {Miyake}}, \bibinfo {author} {\bibfnamefont {G.~A.}\ \bibnamefont
  {Siviloglou}}, \bibinfo {author} {\bibfnamefont {G.}~\bibnamefont {Puentes}},
  \bibinfo {author} {\bibfnamefont {D.~E.}\ \bibnamefont {Pritchard}}, \bibinfo
  {author} {\bibfnamefont {W.}~\bibnamefont {Ketterle}},\ and\ \bibinfo
  {author} {\bibfnamefont {D.~M.}\ \bibnamefont {Weld}},\ }\bibfield  {title}
  {\bibinfo {title} {Bragg scattering as a probe of atomic wave functions and
  quantum phase transitions in optical lattices},\ }\href@noop {} {\bibfield
  {journal} {\bibinfo  {journal} {Physical Review Letters}\ }\textbf {\bibinfo
  {volume} {107}},\ \bibinfo {pages} {175302} (\bibinfo {year}
  {2011})}\BibitemShut {NoStop}%
\bibitem [{\citenamefont {J.~Ma}\ and\ \citenamefont {Nori}(2011)}]{Review}%
  \BibitemOpen
  \bibfield  {author} {\bibinfo {author} {\bibfnamefont {C.~S.}\ \bibnamefont
  {J.~Ma}, \bibfnamefont {X.~Wang}}\ and\ \bibinfo {author} {\bibfnamefont
  {F.}~\bibnamefont {Nori}},\ }\bibfield  {title} {\bibinfo {title} {Quantum
  spin squeezing},\ }\href@noop {} {\bibfield  {journal} {\bibinfo  {journal}
  {Phys. Rep.}\ }\textbf {\bibinfo {volume} {509}},\ \bibinfo {pages} {89}
  (\bibinfo {year} {2011})}\BibitemShut {NoStop}%
\bibitem [{\citenamefont {Sørensen}\ and\ \citenamefont
  {K.Mølmer}(2001)}]{Mol}%
  \BibitemOpen
  \bibfield  {author} {\bibinfo {author} {\bibfnamefont {A.~S.}\ \bibnamefont
  {Sørensen}}\ and\ \bibinfo {author} {\bibnamefont {K.Mølmer}},\ }\bibfield
  {title} {\bibinfo {title} {Entanglement and extreme spin squeezing},\
  }\href@noop {} {\bibfield  {journal} {\bibinfo  {journal} {Phys. Rev. Lett.}\
  }\textbf {\bibinfo {volume} {86}},\ \bibinfo {pages} {4431} (\bibinfo {year}
  {2001})}\BibitemShut {NoStop}%
\bibitem [{Note6()}]{Note6}%
  \BibitemOpen
  \bibinfo {note} {D. J. Wineland, J. J. Bollinger, W. M. Itano, F. L. Moore,
  and D. J. Heinzen, Spin squeezing and reduced quantum noise in spectroscopy,
  Phys. Rev. A {\protect \bf 46}, R6797 (1992).}\BibitemShut {Stop}%
\bibitem [{Note7()}]{Note7}%
  \BibitemOpen
  \bibinfo {note} {P. Hyllus, L. Pezzé, and A. Smerzi, Entanglement and
  Sensitivity in Precision Measurements with States of a Fluctuating Number of
  Particles, Phys. Rev. Lett. {\protect \bf 105}, 120501 (2010).}\BibitemShut
  {Stop}%
\bibitem [{\citenamefont {Toth}(2012)}]{Tot}%
  \BibitemOpen
  \bibfield  {author} {\bibinfo {author} {\bibfnamefont {G.}~\bibnamefont
  {Toth}},\ }\bibfield  {title} {\bibinfo {title} {Multipartite entanglement
  and high-precision metrology},\ }\href@noop {} {\bibfield  {journal}
  {\bibinfo  {journal} {Phys. Rev. A}\ }\textbf {\bibinfo {volume} {85}},\
  \bibinfo {pages} {022322} (\bibinfo {year} {2012})}\BibitemShut {NoStop}%
\end{thebibliography}%

%
%
%
  
\end{document}